\documentclass[letter, amsfonts, amssymb, amsmath, showkeys, nofootinbib, floatfix, 10pt, pre, twocolumn,superscriptaddress]{revtex4-1}
\usepackage[english]{babel}
\usepackage[utf8]{inputenc}
\usepackage[colorinlistoftodos, color=green!40, prependcaption]{todonotes}
\usepackage{amsthm}
\usepackage{mathtools}
\usepackage{physics}
\usepackage{xcolor}
\usepackage{graphicx}
\usepackage[left=23mm,right=13mm,top=35mm,columnsep=15pt]{geometry} 
\usepackage[export]{adjustbox}
\usepackage{placeins}
\usepackage[T1]{fontenc}
\usepackage{lipsum}
\usepackage{csquotes}
\usepackage[caption=false, position=top]{subfig}

\usepackage{siunitx}
\usepackage{float}
\usepackage{dsfont}
\usepackage{amsmath}
\usepackage{booktabs}
\usepackage{tabularx}

\newif\ifdraft

\ifdraft
\def\jrf#1{{\color{orange} [JRF: #1]}}
\def\ms#1{{\color{blue} [MS: #1]}}
\def\hk#1{{\color{green} [HK: #1]}}
\def\hn#1{{\color{red} [HN: #1]}}
\def\jk#1{{\color{pink} [JK: #1]}}
\def\kb#1{{\color{purple} [KB: #1]}}
\def\al#1{{\color{gray} [AL: #1]}}
\else
\def\jrf#1{}
\def\ms#1{}
\def\hk#1{}
\def\hn#1{}
\def\jk#1{}
\def\kb#1{}
\def\al#1{}
\fi

\newcommand{\hhat}{\hat{H}}
\newcommand{\lhat}{\hat{L}}
\newcommand{\htarget}{\hhat_{\mathrm{t}}}
\newcommand{\hinitial}{\hhat_{\mathrm{i}}}
\newcommand{\tfinal}{t_{\mathrm{f}}}

\newcommand{\uj}{u_j}

\newcommand{\xbf}{\mathbf{x}}
\newcommand{\thetabf}{\boldsymbol\theta}
\newcommand{\fsurr}{\Tilde{f}}

\DeclareMathOperator*{\argmax}{arg\,max}

\newcommand{\pauli}[2][]{\hat{\sigma}^{#2}_{#1}}
\newcommand{\spin}[2][]{\hat{S}^{#2}_{#1}}

\newcommand{\tauomega}{\tau_{\Omega}}
\newcommand{\deltainitial}{\Delta_{\mathrm{i}}}
\newcommand{\deltafinal}{\Delta_{\mathrm{f}}}
\newcommand{\hp}{\mathcal{HP}}

\DeclarePairedDelimiter{\nint}\lfloor\rceil

\usepackage[pdftex, pdftitle={Article}, pdfauthor={Author}]{hyperref} 
\begin{document}
\title{Designing Quantum Annealing Schedules using Bayesian Optimization}

\author{Jernej Rudi Fin\v zgar}
    \email[Correspondence email address:\\]{jernej-rudi.finzgar@\{tum.de, bmwgroup.com\}}
    \affiliation{BMW AG, Munich, Germany}
    \affiliation{Technical University Munich, School of CIT, Department of Computer Science, Garching, Germany}
\author{Martin J. A. Schuetz}
    \affiliation{Amazon Quantum Solutions Lab, Seattle, WA, USA}
    \affiliation{AWS Center for Quantum Computing, Pasadena, CA, USA}
\author{J. Kyle Brubaker}
    \affiliation{Amazon Quantum Solutions Lab, Seattle, WA, USA}
\author{Hidetoshi Nishimori}
    \affiliation{International Research Frontiers Initiative, Tokyo Institute of Technology, Shibaura, Minato-ku, Tokyo 108-0023, Japan}
    \affiliation{Graduate School of Information Sciences, Tohoku University, Sendai 980-8579, Japan}
    \affiliation{RIKEN, Interdisciplinary Theoretical and Mathematical Sciences (iTHEMS), Wako, Saitama 351-0198, Japan}
\author{Helmut G. Katzgraber}
    \affiliation{Amazon Quantum Solutions Lab, Seattle, WA, USA}
\date{\today} 

\begin{abstract}

    We propose and analyze the use of Bayesian optimization techniques to design quantum annealing schedules with minimal user and resource requirements. We showcase our scheme with results for two paradigmatic spin models. We find that Bayesian optimization is able to identify schedules resulting in fidelities several orders of magnitude better than standard protocols for both quantum and reverse annealing, as applied to the $p$-spin model. We also show that our scheme can help improve the design of hybrid quantum algorithms for hard combinatorial optimization problems, such as the maximum independent set problem, and illustrate these results via experiments on a neutral atom quantum processor available on Amazon Braket. 
    
\end{abstract}


\maketitle


\section{Introduction} \label{sec:introduction}

\begin{figure*}[htb]
\centering
\includegraphics[width=\textwidth]{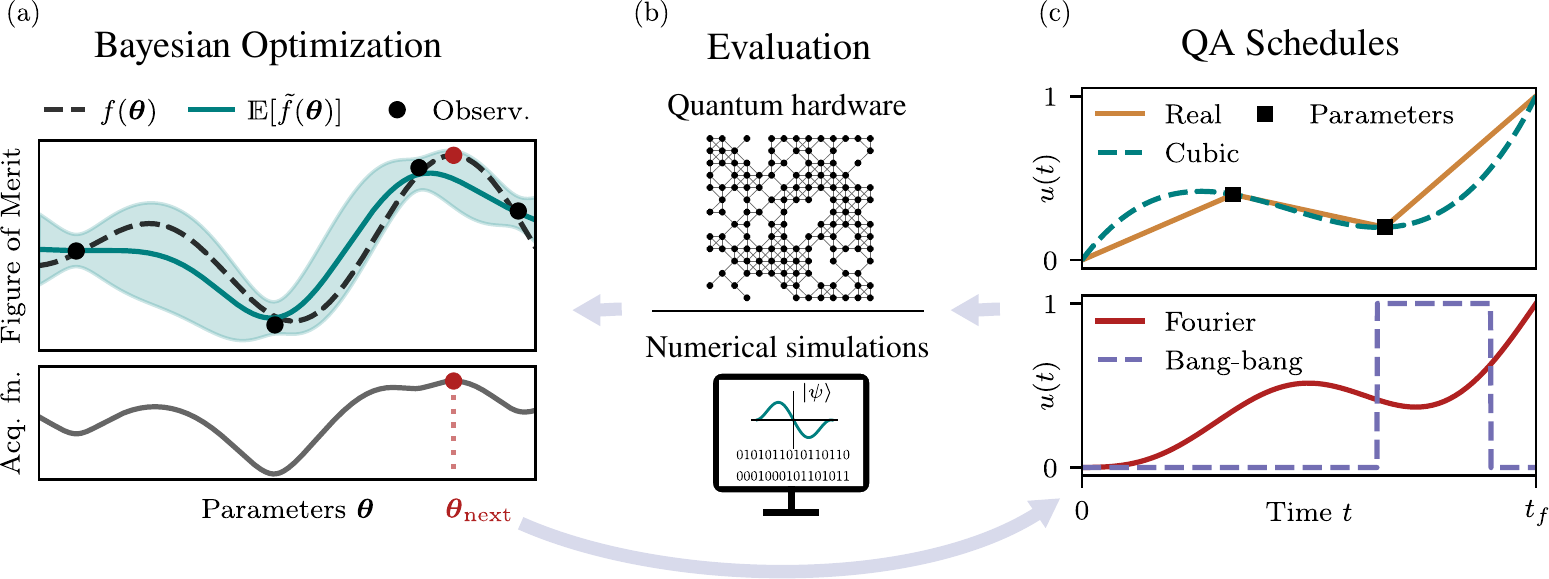}
\caption{Illustration of the proposed schedule optimization scheme. Panel (a) illustrates Bayesian optimization: a surrogate model $\Tilde{f}(\thetabf)$ is built, which represents our belief about the true merit function $f(\thetabf)$. The surrogate model is then used to compute the acquisition function (lower panel in (a)) -- the parameter set for which the acquisition function is maximized is chosen as the next set to probe $\thetabf_{\mathrm{next}}$.
These parameters are then used to specify the schedules $u(t)$, visualized in (c) (with $u(\cdot)$ referring to some relevant Hamiltonian parameter such as, for example, detuning or driving). The schedules are then evaluated in numerical simulations or on actual quantum hardware.
As the outcome can be subject to (e.g., sampling) noise, the obtained values (black circles in (a)) fluctuate slightly around the ground truth (black dashed line in (a)).
The evaluation outcome is then used to update $\Tilde{f}(\thetabf)$, and the cycle can then start anew. We provide further details in the main text.
}
\label{fig:fig1}
\end{figure*}

Over the last few decades, quantum annealing (QA) has emerged as a novel computational paradigm~\cite{Kadowaki1998-wl, Farhi2001-ew}. According to the adiabatic theorem, (ideal) QA has theoretically been shown to be computationally equivalent to the more prevalent circuit model of quantum computing~\cite{Aharonov2004}. However, QA has garnered significant interest mostly due to its potential to encode and solve combinatorial optimization problems~\cite{lucas_ising_2014, hauke_perspectives_2020,Mohseni_2022,Grant_2020, Venegas_Andraca_2018}, with important applications across science and industry. Still, several pitfalls remain -- most notably, the problem of vanishing gaps during the anneal, which results in the run time of QA scaling exponentially for some problems~\cite{znidaric_exponential_2006, farhi2008}. Several extensions have been proposed to overcome this problem, such as leveraging diabatic effects~\cite{crosson_prospects_2021}, counter-diabatic driving~\cite{berry_transitionless_2009, del_Campo_2013, sels_minimizing_2017} and reverse annealing~\cite{ohkuwa_reverse_2018, yamashiro_dynamics_2019}, to name a few.

Recently, variational quantum adiabatic algorithms have been proposed in which a classical optimizer designs QA schedules within a larger hybrid (quantum-classical) computation~\cite{Matsuura_2021, susa_variational_2021, Schiffer_2022, Passarelli_2022,Imoto_2022}. The goal is then to find schedules with the ability to automatically adjust these to the gap structure of the Hamiltonian at hand; for example, by adjusting the annealing speed with which different parameter regions are traversed, or in the form of diabatic paths, which circumvent the need to adhere to the adiabatic speed limit~\cite{crosson_prospects_2021}. Several methods have previously been proposed to find such optimized schedules, including optimal control techniques~\cite{quiroz_robust_2019, brif_exploring_2014, kazhybekova_minimal_2022}, genetic algorithms~\cite{hegde_genetic_2022}, machine learning techniques~\cite{dalgaard_global_2020, yang_optimizing_2020, chen_optimizing_2022, hegde_deep_2022}, as well as several other gradient-based and gradient-free optimizers~\cite{Schiffer_2022, ebadi_quantum_2022}. 

In this work, we consider the use of \emph{Bayesian optimization} (BO) to design high-quality QA schedules with minimal user requirements. BO is a widely used global optimization strategy~\cite{Mockus2011}, especially useful when optimizing (black box) cost functions that are expensive to evaluate, be it in terms of run time or actual monetary cost. This frugal nature of BO makes it an attractive candidate to optimize quantum annealing schedules, especially in the current noisy intermediate scale quantum (NISQ) era, where access to quantum devices is limited and expensive. Previously, BO has already been used to prepare desired states in ultra-cold gases~\cite{mukherjee_preparation_2020, mukherjee_bayesian_2020}, to find optimal angles in the quantum approximate optimization algorithm (QAOA)~\cite{tibaldi_bayesian_2022}, and to improve QA on D-Wave quantum annealers~\cite{pelofske_advanced_2020}.

Here, we investigate the use of BO for preparing ground states (or low-energy states) of two paradigmatic spin models. We first show that simple, hands-off BO schemes (as schematically illustrated in Figure~\ref{fig:fig1}) are able to design QA schedules resulting in fidelities several orders of magnitude better than simple linear schedules. Additionally, we design reverse annealing (RA) schedules for the $p$-spin model and show that BO can identify schedules that successfully circumvent the first-order phase transition. Furthermore, we present numerical evidence that these BO-designed RA schedules are robust against dephasing noise, thereby providing an improvement to previous limitations while using RA~\cite{passarelli_standard_2022}. Finally, we turn our attention to the NP-hard maximum independent set (MIS) problem. In particular, we demonstrate (both in numerical simulations and on actual quantum hardware available on Amazon Braket) that BO can be used to design efficient protocols to solve the MIS problem on neutral atom quantum devices.

The remainder of this paper is organized as follows. In Section~\ref{sec:methods} we introduce Quantum Annealing and Bayesian Optimization as the two essential techniques underlying this work. In Section~\ref{sec:model-systems} we then introduce the model systems on which we test our approach, followed by Section~\ref{sec:results} where we present the results of numerical simulations of the $p$-spin model and results on solving the MIS problem on an array of Rydberg atoms. In Section~\ref{sec:discussion} we discuss the implications of these results and in Section~\ref{sec:outlook} we lay out some of the future research.


\section{Methods} \label{sec:methods}

In this section we outline the main methods used in this paper. First, we briefly introduce QA and define several families of schedule parametrizations as used throughout this manuscript. We then review BO and describe in detail the application thereof in the context of QA schedules.

\subsection{Quantum Annealing}

\subsubsection{General description}

In QA we aim to prepare the ground state of a target Hamiltonian $\htarget$ by exploiting our ability to control the \emph{schedules} $\uj(t)$ of the time-dependent QA Hamiltonian 
\begin{equation}
\label{eq:general-annealing}
    \hhat(t)
    =
    \sum_{j=1}^{n_H} \uj (t) \hat{H}_{j},
\end{equation}
which governs the time evolution of target system. Note that, in general, the Hamiltonian $\hhat(t)$ can be composed of $n_{H}$ individual terms. 

Typically, we begin the procedure with the state of the system $\ket{\psi(0)}$ equal to the ground state of $\hinitial\equiv \hhat(0)$, i.e., we choose the initial Hamiltonian $\hinitial$ such that its ground state is easy to prepare. We then evolve the state with the time-dependent Hamiltonian Eq.~\eqref{eq:general-annealing} for a time $\tfinal$ such that ultimately $\hhat(\tfinal)=\htarget$. If the rate of change of the Hamiltonian is slow enough compared to the minimum of the spectral gap $\Delta(t)$ between the instantaneous ground and first excited states of $\hhat(t)$, then the success of QA is guaranteed by the adiabatic theorem; see, for example~\cite{albash_adiabatic_2018}. If the requirement of adiabaticity is abandoned, one may (for example) leverage \emph{diabatic} effects, which can potentially yield shortcuts to the ground state of $\htarget$~\cite{crosson_prospects_2021, GuryOdelin2019}. However, such paths might be difficult to find in practice. Finally, one can additionally forgo the requirement that $\hhat(\tfinal)=\htarget$; this has been explored in the context of QAOA-inspired so-called \emph{bang-bang} schedules (see Sec.~\ref{sec:schedule-parametrizations}). 

\subsubsection{Schedule parametrizations}
\label{sec:schedule-parametrizations}

Here, we describe the schedule parametrizations used in this work, as summarized in Table~\ref{tab:parametrizations}. For convenience, we first focus on the first five parametrizations and address the bang-bang protocol individually at the end of this section. All of these five Ansaetze are designed such that they fulfill the boundary conditions $u(0)=0$ and $u(\tfinal)=1$, and can be used as basic building blocks for the design of more sophisticated schedules by transforming or combining them according to the required boundary conditions, e.g., $u(t)\rightarrow 1-u(t)$.
Thus, one can use these parametrizations to construct the desired QA Hamiltonian Eq.~\eqref{eq:general-annealing}.

\begin{table}[htb]
\begin{tabularx}{\columnwidth}{@{}llc@{}}
\toprule[1.0pt]
\textbf{Schedule} & \textbf{Function} $u(t)$ & $\theta_j\in$\\ \midrule[0.2pt]
Linear & $t/\tfinal$ &  \multicolumn{1}{c}{/}   \\
Real & $\mathrm{LinInterp}(\left\{0, \theta_1, ..., \theta_{n}, 1\right\})(t)$ & $[\frac{j-\zeta}{n+1}, \frac{j+\zeta}{n+1}]$ \\
Cubic & $\mathrm{CubInterp}(\left\{0, \theta_1, ..., \theta_{n}, 1\right\})(t)$ & $[\frac{j-\zeta}{n+1}, \frac{j+\zeta}{n+1}]$ \\
Low-pass & $\mathrm{LowPass}(\left\{0, \theta_1, ..., \theta_{n}, 1\right\})(t)$ & $[\frac{j-\zeta}{n+1}, \frac{j+\zeta}{n+1}]$ \\
Fourier & $t/\tfinal + \sum_{j=1}^n \theta_j\sin\left(j\pi t/\tfinal\right)$ & $[-\frac{1}{j}, \frac{1}{j}]$\\
Bang-bang & $\frac{\theta_1}{\tfinal}[H(t-t_1) - H(t_2 - t)]$ & $[0, 2\pi]$ \\
\bottomrule[1.0pt]
\end{tabularx}
\caption{The functional form of the various schedule parametrizations used in this work; see Fig.~\ref{fig:fig1}c for a visualization. Here $H(\cdot)$ denotes the standard Heaviside step function. The rightmost column gives the parameter bounds for the individual parameters.}
\label{tab:parametrizations}
\end{table}

The \emph{real}, \emph{cubic}, and \emph{low-pass} parametrizations all take a collection of $n$ points $\theta_j$ as inputs. These specify the value of the schedule at equidistant points in time $\theta_j=u(j\tfinal/(n + 1))$, where $j=1,...,n$. In addition $u(0)=0$ and $u(\tfinal)=1$. In the real (cubic) parametrization linear (cubic) interpolation is then used to obtain the intermediate values, resulting in a piecewise linear (cubic) schedule. For the low-pass parametrization, the real schedule is additionally passed through a low-pass filter, which smoothens the piece-wise linear schedule (see Fig.~\ref{fig:rydberg-schedules} for an example). This has been suggested previously to ensure that the schedules comply with the limitations of the neutral atom experimental platform, and to mitigate potential artifacts (e.g., excitations) at the points where the original schedule is not differentiable~\cite{ebadi_quantum_2022}. The degree to which the schedules are allowed to deviate from the linear schedule is controlled by the $\zeta$ parameter (see right column of Table~\ref{tab:parametrizations}). We use $\zeta=2$ unless stated otherwise.

In the case of the \emph{Fourier} parametrization, sinusoidal deviations from the linear schedule are considered, with $\theta_j$ setting the magnitude of the term with frequency $\nu_j = j/(2\tfinal)$. As we found in practice that high-frequency components are less important, we decrease their allowed magnitude with increasing $j$, thus reducing the search space for the optimizer.

Finally, we turn our attention to the \emph{bang-bang} schedule, which is qualitatively different from the others, as it is inspired by QAOA-type algorithms designed for gate-based architectures. As shown in Fig.~\ref{fig:fig1}c, it consists of constant-magnitude discrete pulses. For the experiments documented in this manuscript, it is sufficient to consider pulses that attain the nonzero value of $\frac{2 \theta_1}{\tfinal}$ in the intervals $[0, \tfinal/2]$ and $[\tfinal/2, \tfinal]$ (i.e., either in the first half or the second half of the protocol), corresponding to setting the parameters $(t_1, t_2)$ in Table~\ref{tab:parametrizations} to either $(0, \tfinal/2)$ or $(\tfinal/2, \tfinal)$, respectively.
Thus, the parameter $\theta_1$ directly corresponds to the area of the pulse.
We only fix the duration of the pulses for convenience. However, in principle, other parametrizations are possible. Consequently, evolving a system with the Hamiltonian $u(t) \hhat$, where $\hhat$ is constant in time, is equivalent to applying the unitary $U(\theta_1)=\mathrm{exp}(-i\hhat \theta_1)$, thus making the connection to QAOA immediately evident; see, for example, Ref.~\cite{qaoa-farhi}.

\subsection{Bayesian Optimization}

\subsubsection{Preliminaries}

Let us  review how BO can be used to optimize a scalar objective function 
$f:\mathcal{D}_{\theta} \rightarrow\mathds{R}$, where $\mathcal{D}_{\theta}\subset \mathds{R}^n$ is the parameter domain. In the most standard variant of BO, Gaussian Processes (GPs) are used to build a surrogate model $\fsurr(\thetabf)$ of the optimization objective. We say that a function $\fsurr(\thetabf)$ is distributed as a GP $\fsurr(\thetabf)\sim\mathcal{GP}\left[\vec{\mu}(\thetabf), k(\thetabf, \thetabf')\right]$, where $\vec{\mu}(\thetabf)$ is the mean and $k(\thetabf, \thetabf')$ is the kernel function, when for every finite subset of points $\Theta=\left\{\thetabf_1, ..., \thetabf_m\right\}$ it holds that $\fsurr(\Theta)\sim \mathcal{N}(\vec{\mu}, \Sigma)$. Here, $\mathcal{N}(\vec{\mu}, \Sigma)$ is a multivariate normal distribution with an $m$-dimensional mean vector $(\vec{\mu})_i=\mu(\thetabf_i)$ and an $m\times m$ covariance matrix $\Sigma_{ij}=k(\thetabf_i, \thetabf_j)$. In other words, $\vec{\mu}$ contains our belief about the mean values of $\fsurr$ evaluated on points in $\Theta$. Similarly, $\Sigma$ holds the information about the correlation between the values of $\fsurr$ on $\Theta$.

We can incorporate our prior beliefs about the objective function through the choice of a kernel function $k$. In our case, we choose a Mat{\'e}rn $5/2$ kernel, which favors continuous functions~\cite{Rasmussen2005-yw}. The chosen kernel is isotropic, meaning that it only depends on the Euclidean distance $d:=d(\thetabf, \thetabf')$ between two points as
$$k(\thetabf, \thetabf'; \ell)
=
\left(
1 + \frac{\sqrt{5}d}{\ell} + \frac{5d^2}{3\ell^2}
\right)
\exp\left(-\frac{\sqrt{5}d}{\ell}\right),
$$
where the length scale $\ell$ determines how quickly the correlations between values at two different points decay as a function of the Euclidean distance $d$ between them.

For instance, if one is interested in the values of the model function $\fsurr$ at a set of points $\Theta$, and one assumes that one already knows the values for some set of observations $\Theta_{\mathrm{obs}}\subset \Theta$. We can condition the probability distribution of $\fsurr$ on the observations $f(\Theta_{\mathrm{obs}})$. A key feature of the normal distribution is that it is closed under conditioning, meaning that the resulting distribution on the remaining points of interest $\Theta_{\mathrm{rem}} = \Theta\setminus \Theta_{\mathrm{obs}}$ is also normally distributed as $\fsurr(\Theta_\mathrm{rem})=\mathcal{N}(\vec{\mu}_{\mathrm{c}}, \Sigma_{\mathrm{c}})$. Here $\vec{\mu}_{\mathrm{c}}$ and $\Sigma_{\mathrm{c}}$ are the conditioned mean vector and covariance matrix, respectively, which can be obtained by elementary matrix algebra (see, e.g., Ref.~\cite{Rasmussen2005-yw}). The straightforward way in which observations can be included into the model is one of the main reasons why GPs are typically used as the surrogate model. Another benefit of BO is that uncertainty about the observations can be easily incorporated into the model, if we assume that $f(\thetabf)$ is a normal random variable centered at the ground truth value, with a standard deviation $\sigma_{\mathrm{obs}}$. This can be useful, for example, to incorporate statistical noise resulting from a finite number of shots obtained from a quantum device. For a visualization see Fig.~\ref{fig:fig1}a, where the observations (black dots) deviate slightly from the ground truth (black dashed line). The updated GP can then be used to determine the next point to probe, which is selected to be the point where the so-called \emph{acquisition function} $\varphi(\thetabf)$ attains its maximal value $\thetabf_{\mathrm{next}} = \argmax\varphi(\thetabf)$. This is shown in the bottom of Fig.~\ref{fig:fig1}a.

Several heuristics for $\varphi$ are used in the literature~\cite{acquisition-functions}; in this work, we use the upper confidence bound~\cite{auer2002using}
\begin{equation}
\label{eq:acquistion-function}
\varphi(\thetabf; \kappa) 
=
\mu(\thetabf) + \kappa\cdot\sigma(\thetabf).
\end{equation}
Here $\mu(\thetabf)=\mathds{E}[\fsurr(\thetabf)]$ is the mean value of $\fsurr(\thetabf)$ (indicated by the teal (solid) line in Fig.~\ref{fig:fig1}a) and $\sigma(\thetabf)=(\mathds{E}[\fsurr(\thetabf)^2] - \mu(\thetabf)^2)^{1/2}$ is its standard deviation (light teal shading in Fig.~\ref{fig:fig1}a). The hyperparameter $\kappa$ sets the relative importance of the mean value of $\fsurr(\thetabf)$ and its standard deviation and thus controls the amount of exploration permitted. In this work we decrease the value of $\kappa$ as the optimization progresses. Intuitively, we want the algorithm to explore more initially and leverage the acquired information about the optimization landscape at later stages, thus gradually switching from exploration to exploitation.

\subsubsection{BO Pipeline}
\label{sec:bo-pipeline}

Let us now provide some more detail about the end-to-end pipeline---illustrated in Fig.~\ref{fig:fig1}---used to produce the results shown in this paper. At the beginning of the process, a schedule Ansatz (or a set thereof) must be specified. This schedule template takes in a set of parameters, which are elements of bounded intervals. We then need to pass the bounds of these parameters to the optimizer to limit the domain on which a surrogate model of the objective function is constructed. Additionally, a figure of merit (FoM), e.g., expected energy, must be specified, which can be obtained by evaluating the schedule at a particular value of the parameters. The evaluation can then be carried out either in numerical simulations or on an actual quantum device. As we use a finite sample of size $N_{\mathrm{shots}}$ to estimate the FoM, we scale the standard deviation of the observations as $\sigma_{\mathrm{obs}}\propto N_{\mathrm{shots}}^{-1/2}$.

Unless specified otherwise, our BO pipeline then proceeds as follows. We initialize the optimizer by probing the linear schedule, i.e., by evaluating the parameters corresponding to the linear schedule for each parametrization. Note that for the bang-bang parametrization, where this is not possible, we start with random parameters. We then evaluate nine additional random parameter sets, after which the next points are chosen by the maximum of the acquisition function Eq.~\eqref{eq:acquistion-function}, starting with $\kappa=2$ (see Fig.~\ref{fig:fig1}a). This is then repeated for a total of 50 iterations. As we would like the optimizer to focus on exploiting the information it has acquired during the initial exploration, we specify a decay schedule for $\kappa$. After 25 iterations, we begin decreasing $\kappa$ as a geometric sequence, such that its final value is $\kappa=0.01$. In the BO implementation we use in this study~\cite{bo-python}, additionally, hyperparameters (such as the kernel length scale $\ell$) of the kernel function are progressively tuned to improve the quality of our surrogate for the objective, making this particular implementation powerful out-of-the-box, without any need for hyperparameter tuning.


\section{Benchmark systems}
\label{sec:model-systems}

In what follows, we introduce the two paradigmatic model systems studied in this work -- quantum annealing (QA) and reverse annealing (RA) of the $p$-spin model, and a variational adiabatic protocol for finding the maximum independent set (MIS) of a graph using a neutral atom quantum processor based on Rydberg atoms.

\subsection{\texorpdfstring{$p$}--spin model}
\label{sec:pspin-model}
The $p$-spin model is a well-studied spin system with all-to-all $p$-body terms. Using the total spin operators 
\begin{equation}
    \spin{x, z}:=\sum_{i=1}^{N} \pauli[i]{x,z},
\end{equation}
where $\pauli[i]{x}$ ($\pauli[i]{z}$) is the usual Pauli-X (Pauli-Z) matrix acting on site $i$, we can write the $p$-spin Hamiltonian for $N$ spins as
\begin{equation}
\label{eq:pspin-hamiltonian}
    \htarget 
    =
    -N\left(N^{-1}\spin{z}\right)^p.
\end{equation}
For $p$ odd the unique ground state of the $p$-spin Hamiltonian is $\ket{\uparrow}^{\otimes N}$, where $\pauli{z}\ket{\uparrow}=+1 \ket{\uparrow}$. Here, we focus on the case where $p=3$, as higher values $p\geq 4$ exhibit qualitatively similar behavior~\cite{yamashiro_dynamics_2019}. We use $\hbar=1$ throughout this section.

\subsubsection{Quantum Annealing}

In the case of QA, we start the protocol in the ground state of the transverse field Hamiltonian
\begin{equation}
\label{eq:vtf-hamiltonian}
    \hhat_{\mathrm{TF}} 
    = 
    -\Gamma\spin{x},
\end{equation}
where $\Gamma$ is the strength of the transverse field. The initial ground state is then just $\ket{+}^{\otimes N}$, where $\ket{+}:=2^{-1/2}(\ket{\uparrow} + \ket{\downarrow})$ is the $+1$ eigenstate of $\pauli{x}$. We then evolve the system via
\begin{equation}
\label{eq:pspin-qa}
    \hhat(t)
    =
    s(t)\hhat_{\mathrm{TF}}
    +
    \left[1-s(t)\right]\htarget,
\end{equation}
where we demand $s(\tfinal)= 1 - s(0)= 0$. Here we have implicitly restricted ourselves to only considering parametrizations where the control functions are \emph{dependent}. It is particularly convenient to use such a parametrization, as fewer parameters are required. Hence, we employ this Ansatz for QA in the $p$-spin model throughout the main text and defer results with \emph{independent} controls to Appendix~\ref{app:independent-controls}.

It has been shown, by means of a static~\cite{ohkuwa_reverse_2018} and a dynamic~\cite{yamashiro_dynamics_2019} analysis, that the system governed by the Hamiltonian in Eq.~\eqref{eq:pspin-qa} exhibits a first-order phase transition as $s(t)$ is varied between $0$ and $1$. A first-order phase transition is typically (but not exclusively, see e.g.~\cite{Laumann2012, Tsuda2013}) associated with an exponentially small (in the size of the system $N$) gap between the ground state and the first excited state of the instantaneous Hamiltonian~\cite{Schtzhold2006, Amin2009}. Consequently, the adiabatic timescale $\tfinal$ likewise suffers from exponential scaling.

One possibility to circumvent the exponentially long time scale was proposed in Ref.~\cite{wauters_polynomial_2020}, where QAOA circuits for preparing the ground state of the $p$-spin Hamiltonian are proposed, which can be turned into constant time annealing schedules~\cite{garcia-pintos_lower_2022}. These consist of two constant valued pulses; for $t\in[0,\tfinal/2)$ with $\htarget$ and subsequently for $t\in(\tfinal/2, \tfinal]$ with $\hhat_{\mathrm{TF}}$. We use the bang-bang schedules defined in Sec.~\ref{sec:schedule-parametrizations} and learn the parameters with BO. Additionally, several other potential modifications to the basic protocol in Eq.~\ref{eq:pspin-qa} have been proposed in the literature~\cite{susa_exponential_2018, koh_reduction_2020, seki_quantum_2012, durkin_quantum_2019, passarelli_improving_2019, passarelli_counterdiabatic_2020}. Here, we focus on Reverse Annealing (RA)~\cite{ohkuwa_reverse_2018,yamashiro_dynamics_2019}.

\begin{figure}[htb]
    \centering
    \includegraphics{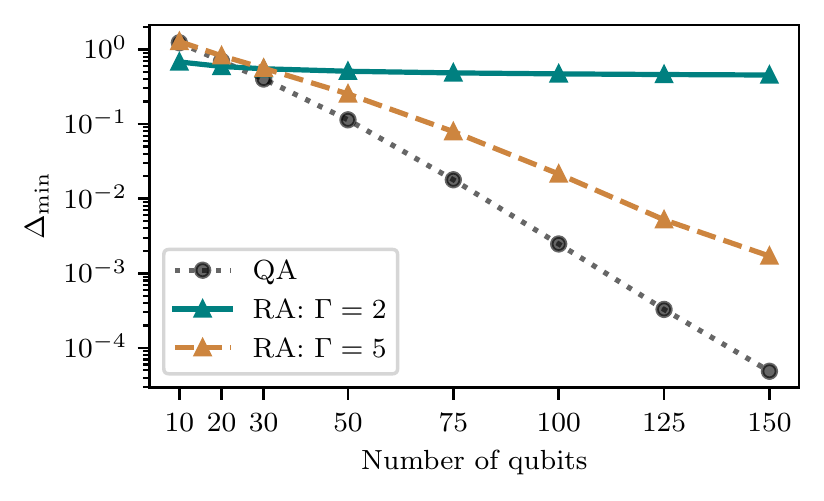}
    \caption{Minimal spectral gap $\Delta_{\min}$ for the linear protocol $s(t)=t/\tfinal$ for QA and for the linear path $s(t)=\lambda(t)=t/\tfinal$ in RA. The results indicate that the exponential closing of the minimal spectral gap in QA can only be circumvented in RA if $\Gamma$ is chosen correctly. RA results shown are for $c=0.8$.
    }
    \label{fig:pspin-gaps}
\end{figure}

\subsubsection{Reverse Annealing}

In RA the system is initialized in a classical state, corresponding to an approximate solution of the target problem, which we have obtained by e.g., a classical approximation algorithm.
Here, we consider states $\ket{\psi_c}$ that have a certain fraction $c\in[0,1]$ of spins aligned with the ground state $\ket{\uparrow\uparrow\dots\uparrow}$ of the $p$-spin Hamiltonian.
Due to the permutation symmetry of the Hamiltonians in Eq.~\eqref{eq:pspin-hamiltonian} and Eq.~\eqref{eq:vtf-hamiltonian}, we can without loss of generality choose a configuration where the first $N_c:=\nint{Nc}$ spins are aligned with the ground state of $\htarget$ and the rest are antialigned as
\begin{equation}
    \ket{\psi_c}
    =
    \mid\hspace{-0.1cm}\underbrace{
    \uparrow\uparrow\dots\uparrow
    }_{N_c}
    \,\rangle_1
    \otimes
    \mid\hspace{-0.1cm}\underbrace{
    \downarrow\downarrow\dots\downarrow
    }_{N - N_c}
    \,\rangle_2.
\end{equation}
This is the ground state of
\begin{equation}
    \hhat_{\mathrm{init}}
    = 
    -\spin[1]{z}
    +
    \spin[2]{z},
\end{equation}
where $\spin[j]{z}$ is the total $z$ projection of the spin in the aligned ($j=1$) and antialigned ($j=2$) sectors, comprising the first $N_c$ and the remaining $N-N_c$ spins, respectively.
We can then define the RA Hamiltonian as
\begin{equation}
\label{eq:ra-hamiltonian}
\begin{split}
    H(t)
    &=
    \left[1-s(t)][1-\lambda(t)\right] H_{\mathrm{init}} 
    +
    s(t)\htarget \\
    &+
    \left[1-s(t)\right]\lambda(t)H_{\mathrm{TF}} 
\end{split}
\end{equation}
with $s(0)=\lambda(0)=0$ and $s(\tfinal)=\lambda(\tfinal)=1$. Note that for $\lambda(t)=1$ we recover standard QA as defined in Eq.~\eqref{eq:pspin-qa}.

It has been shown that for sufficiently good initial approximations (i.e., for $c$ sufficiently close to $1$), the first-order phase transition in the $p$-spin model can be avoided by RA~\cite{ohkuwa_reverse_2018}. This has been leveraged to design polynomial-time adiabatic protocols to prepare the ground state of the $p$-spin model~\cite{yamashiro_dynamics_2019}. However, as the gap structure of the RA Hamiltonian Eq.~\eqref{eq:ra-hamiltonian} is \emph{a priori} unknown, such a speedup is subject to finding the correct parameters (e.g., the transverse field strength $\Gamma$, see Fig.~\ref{fig:pspin-gaps}), especially if we consider exclusively linear schedules for $\lambda(t)$ and $s(t)$ (see background in Fig.~\ref{fig:ra-paths}). Hence, finding a way to design $s(t)$ and $\lambda(t)$ to successfully navigate the $\Delta(s, \lambda)$ landscape is essential to harnessing the full speedup obtainable in RA.

An essential observation is that $\forall (s,\lambda)$ the RA Hamiltonian Eq.~\eqref{eq:ra-hamiltonian} commutes with the squared total spin operator $\mathbf{S}_j^2$ in both sectors $j=1,2$.  Because the initial state belongs to the largest eigenvalues of $\mathbf{S}_j^2$ in both sectors, the evolution is constrained to this subspace. Therefore, only a Hilbert space of dimension $(N_c+ 1)(N - N_c+ 1)$ needs to be considered, which makes larger system sizes accessible by numerical simulation. Note that this also holds for QA, with a Hilbert space dimension $N + 1$ obtained by setting $c=0$. In this work, numerical simulations are performed using the \texttt{QuTiP} library~\cite{Johansson_2012, Johansson2013}.


\subsubsection{Noise in the \texorpdfstring{$p$}--spin model}
\label{sec:pspin-noise}

It has been shown that RA loses some of its computational advantage over QA when dephasing noise is considered~\cite{passarelli_standard_2022}. To analyze this statement in more detail within our approach, we set up simulations of the $p$-spin model in the mathematical framework of the Adiabatic Master Equation (AME)~\cite{albash_quantum_2012}:
\begin{equation}
\label{eq:ame}
    \partial_t \rho(t)
    =
    -i\left[H(t)+H_{\mathrm{LS}}(t), \rho(t)\right]
    +
    \mathcal{D}[\rho(t)].
\end{equation}
Here, $H_{\mathrm{LS}}$ is the Lamb-Shift Hamiltonian, and $\mathcal{D}$ is the dissipator -- for details we refer the reader to Appendix \ref{app:ame}. To be able to simulate larger systems, we limit ourselves to the study of the AME with \emph{collective} dephasing, which does not break permutation symmetry but was shown to hinder the performance of RA~\cite{passarelli_standard_2022}. We employ the \texttt{HOQST} library to perform numerical simulations of the AME~\cite{chen_hamiltonian_2022}.

\subsection{Maximum Independent Set and Neutral Atom Quantum Computers}

\subsubsection{Maximum Independent Set Problem}

Given a graph $G=(V,E)$ with a set of vertices $V$ and an edge set $E$ an independent set is a subset of vertices $S\subseteq V$ such that no pair of vertices $v_i, v_j\in S$ in the subset are connected by an edge $(v_i, v_j)\notin E,\, \forall v_i, v_j \in S$. The largest independent set is called the maximum independent set (MIS). Finding the MIS is known to be NP-hard for general graphs~\cite{Karp1972}, making the existence of an efficient (i.e., polynomial time) algorithm unlikely. However, applications of the MIS problem can be found across a broad range of industries~\cite{wurtz_industry_2022}. Here, we focus our attention on solving the MIS problem on a family of graphs called unit-disk graphs (dubbed UDGs in the following). These are generated by positioning a collection of vertices in the plane and only connecting those that are less than a certain distance $R_{\mathrm{d}}$ apart. Despite being a special family of graphs, finding the MIS of a general UDG is still an NP-Hard problem~\cite{pichler_quantum_2018}. For details on the graph generation in this work, refer to Appendix \ref{app:mis}.

\subsubsection{MIS with Rydberg Atom Arrays}
\label{sec:mis-rydberg}

The MIS problem on UDGs naturally emerges in the context of neutral atom quantum computers based on Rydberg atom arrays~\cite{ebadi_quantum_2022, pichler_quantum_2018}. In what follows, we briefly summarize the aspects of neutral atom quantum computation relevant to the present study -- for a comprehensive review, we refer the reader to Ref.~\cite{Henriet_2020}.

\begin{figure}[htb]
    \centering
    \includegraphics{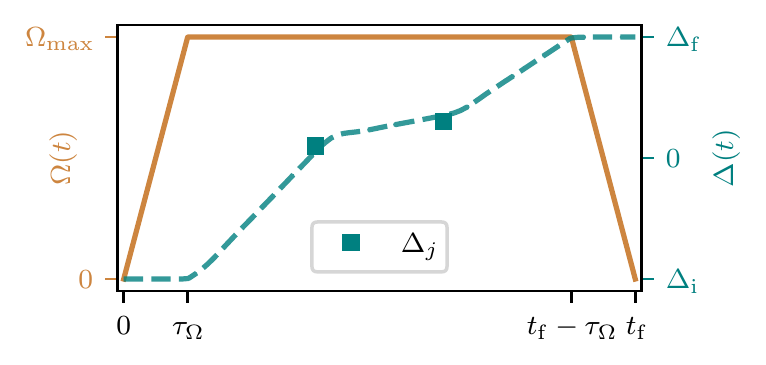}
    \caption{Visualizations of the schedules for the Rabi frequency $\Omega(t)$ and the laser detuning $\Delta(t)$. The $\Omega$ schedule (solid line) takes in its maximal value $\Omega_{\max}$, and the ramp-up time $\tau_{\Omega}$ as parameters. The detuning schedule is specified by a collection of (in this case 2) equally spaced points $\Delta_j$. A linear interpolation between these points is then passed into a low-pass filter, yielding the dashed teal schedule depicted above.
    }
    \label{fig:rydberg-schedules}
\end{figure}

The dynamics of neutral atom quantum processors are governed by the following Hamiltonian $\hat{H}(t)=\hhat_{\mathrm{dr}} + \hhat_{\mathrm{cost}}$ with 
\begin{equation}
\label{eq:rydberg-hamiltonian}
\begin{aligned}
    \hhat_{\mathrm{dr}}
    &=
    \frac{\hbar}{2} \Omega(t) \sum_i
    \pauli[i]{x}
    \\
    \hhat_{\mathrm{cost}}
    &=
    -\hbar \Delta(t) \sum_i \hat{n}_i
    +
    \sum_{i<j} V_{i j} \hat{n}_i \hat{n}_j,
\end{aligned}
\end{equation}
where $\Omega$ is the Rabi frequency, $\Delta$ is the laser detuning, and $V_{ij}=C_6\lVert \xbf_i - \xbf_j \lVert_2^{-6}$, with $C_6$ the van der Waals coefficient, which depends on the particular atomic species used. The number operator $\hat{n}_i:=\mathds{1}_2-\pauli[i]{z}$ counts the number of Rydberg excitations at site $i$.

A crucial feature of the Hamiltonian in Eq.~\eqref{eq:rydberg-hamiltonian} is the so-called Rydberg blockade phenomenon, in which two atoms cannot simultaneously be in the (excited) Rydberg state $\ket{1}$ if their spatial separation is smaller than the Rydberg blockade radius $R_{\mathrm{b}} \equiv \left(
C_6/
\hbar\Omega
\right)^{1/6}$~\cite{ebadi_quantum_2022}.
Therefore, if one sets the UDG length scale as $R_{\mathrm{d}}=R_{\mathrm{b}}$, ground states of $\hhat_\mathrm{cost}$ naturally obey the independence constraint on UDGs if one associates atoms that are in the $\ket{1}$ with membership in the MIS~\cite{pichler_quantum_2018}. If we furthermore set $0 < \Delta < V_{ij}$, in the ground state of $H_\mathrm{cost}$, the number of excitations will be maximized, while not violating the independence constraint, thus corresponding to solutions to the original MIS problem. We thus start the adiabatic protocol with $\Delta < 0$, such that the ground state is $\ket{0}^{\otimes |V|}$ and then (slowly) increase the detuning until some final positive value, as shown in Fig.~\ref{fig:rydberg-schedules}, such that the ground states of the final Hamiltonian correspond to the solutions of the MIS problem.

In our experiments, we run the protocols identified by the BO algorithm and then use the top $50\%$ of the bit strings in terms of the following MIS energy (cost) function
\begin{equation}
\label{eq:mis-cf}
    H(\xbf) 
    =
    -\sum_{i=1}^{|V|} x_i
    +
    \alpha \sum_{(i, j)\in E} x_i x_j,
\end{equation}
where $x_i$ is the $i$-th component of the bit string $\xbf\in\{0,1\}^{|V|}$, which indicates if the $i$-th vertex is present in the independent set ($x_i=1$). We have found the value $\alpha=1.2$ to work well in practice, and thus use it throughout this manuscript unless specified otherwise.


\begin{figure*}[ht]
\centering
\subfloat[]{%
\includegraphics{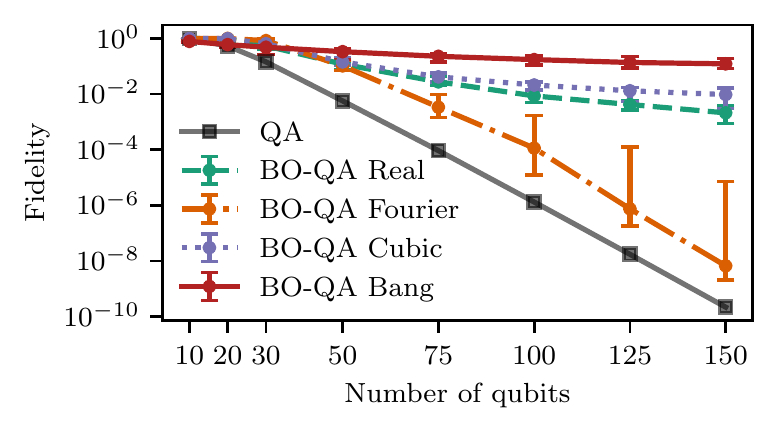}%
\label{fig:qad-scaling}%
}
\qquad
\subfloat[]{%
\includegraphics{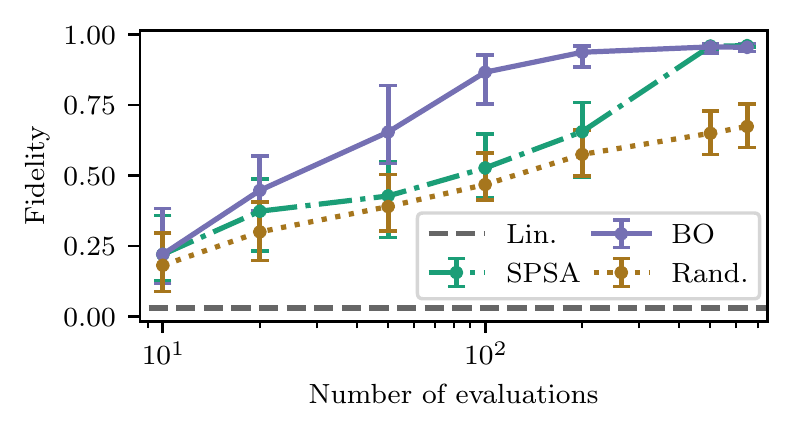}%
\label{fig:iteration-number-scaling}%
}
\caption{(a) System size scaling of the fidelity for BO-QA for the $p$-spin model, with different parametrizations of control functions (see Sec.~\ref{sec:schedule-parametrizations} for details). (b) Scaling of the fidelity obtained by quantum annealing schedules designed with a varying number of iterations of BO and SPSA. For comparison, we also show the fidelities obtained by evaluating the same number of random parameter sets. A real-space parametrization with $4$ parameters was used for a $p$-spin model with $N=15$ spins and $\tfinal=3$. For reference, the results of the linear schedule are shown, which achieves a fidelity of $\sim 0.03$ only. All strategies were trained using fidelity as the figure of merit for the optimizer.}
\label{fig:pspin-fig1}
\end{figure*}

\section{Results} 
\label{sec:results}

In what follows, we present results for BO-designed annealing schedules. First, we present results based on numerical simulations of the $p$-spin model, which allow us to study (relatively) large system sizes of a paradigmatic model that exhibits a first-order phase transition. Then, we turn our attention to the MIS problem on neutral atom quantum computers. Here, we perform numerical (classical) simulations for small graphs and compare our results with results obtained from the QuEra Aquila device~\footnote{\href{https://www.quera.com/aquila}{QuEra Aquila} device was accessed through \href{https://aws.amazon.com/braket/quantum-computers/quera/}{Amazon Braket}.}. Finally, we implement and execute our end-to-end optimization pipeline on the QuEra device.

As BO is a stochastic optimizer, averages over several runs are required to assess its (average) performance. We report median values of the performance metrics, with the error bars denoting the upper and lower quartiles. In all experiments at least $80$ repetitions have been performed.


\subsection{\texorpdfstring{$p$}--spin model}

We first analyze the $p$-spin model in (classical) numerical simulations and set the strength of the transverse field $\Gamma=5$ and use $p=3$, unless stated otherwise. We present results for both QA and RA schedules and study the effects of noise. 

We begin with a characterization of BO and different annealing schedules by considering the system size scaling of the ground-state fidelities obtained at the end of the protocol for the case where ground state fidelity is likewise used as the figure of merit for the optimizer. While this scenario might not be realistic for many real-world situations because the ground state is typically not known a priori, it provides the easiest setup for the optimizer to learn, because the feedback is obtained directly on the relevant FoM. In Fig.~\ref{fig:qad-scaling}, the fidelity for different parametrizations is shown. All parametrizations depend on four parameters, apart from the bang-bang protocol, which uses just two.

\subsubsection{Quantum Annealing}

The fact that the system undergoes a first-order phase transition manifests itself in the dramatic decrease in fidelity for the na\" ive linear schedule. The fidelity can be improved by several orders of magnitude (when the number of variables increases) using BO, for all of the parametrizations considered. The best results are obtained when the bang-bang parametrization is used -- in line with previous results showing that it is possible to prepare the ground state of the $p$-spin model with an optimal $\tfinal\sim\mathcal{O}(1)$ schedule~\cite{wauters_polynomial_2020, garcia-pintos_lower_2022}. However, similarly, our results based on the real and cubic parametrizations are able to improve dramatically on the linear QA schedule as well.

\begin{figure}[htb]
    \centering
    \includegraphics{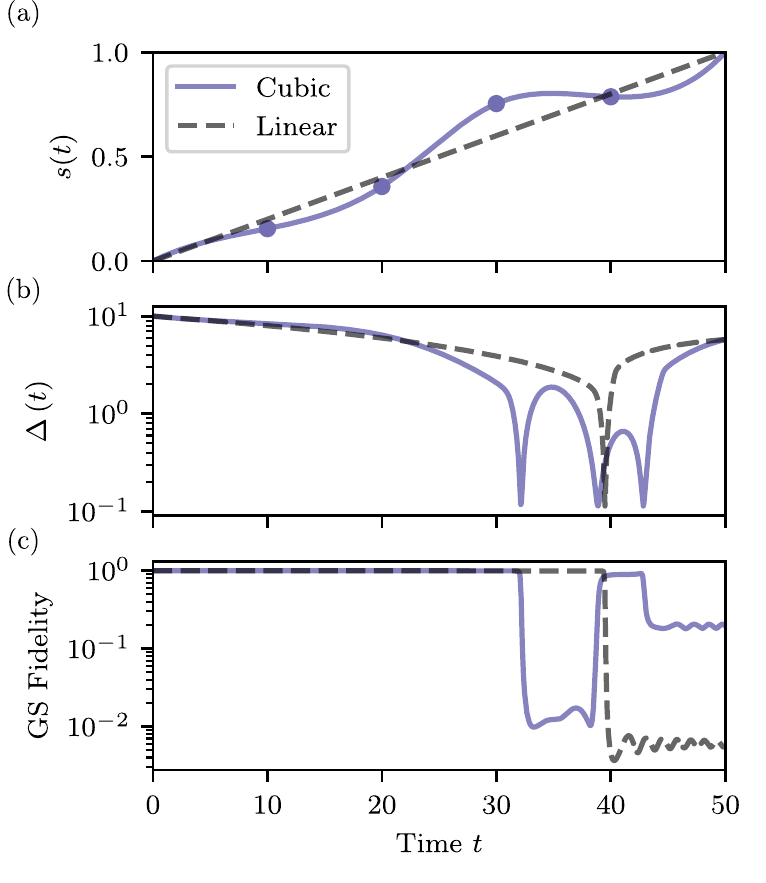}
    \caption{(a) Illustration of a prototypical cubic schedule, which is obtained by cubic interpolation between the points (purple dots) determined by BO, and the naive linear (baseline) schedule. The cubic schedule is specified by the purple dots. 
    (b) Instantaneous spectral gap $\Delta(t)$ for both schedules.
    (c) Instantaneous ground-state fidelity for both schedules, indicating that diabatic mechanisms are facilitating the improvements of the BO schedule.
    Results refer to the $p$-spin model with $p=3$, $N=50$ spins and $\tfinal=50$. 
    }
    \label{fig:qa-gap-paths}
\end{figure}

As mentioned previously, a key reason to consider BO for the optimization of annealing schedules lies in its efficient use of resources (i.e., queries of the cost function). We showcase this feature of BO explicitly through complementary benchmark results obtained with the Simultaneous Perturbation Stochastic Approximation (SPSA) algorithm~\cite{spsa}, a common choice when optimizing variational quantum algorithms (see, for example, Refs.~\cite{ebadi_quantum_2022, Kandala_2017, cerezo_variational_2021}). Specifically, for QA in the $p$-spin model with $N=15$ spins and $\tfinal=3$, we compare the performance of SPSA and BO in terms of the fidelity achieved as the number of allowed evaluations is increased (see Fig.~\ref{fig:iteration-number-scaling}). Here, hyperparameter optimization was performed for both SPSA and BO to allow for a fair comparison. For reference, the same number of random parameter evaluations is shown. In the intermediate regime (with $\sim 50$ -- $200$ evaluations), BO is able to outperform SPSA-based schedules, in line with previous studies using BO to control quantum systems~\cite{mukherjee_preparation_2020}. As we increase the number of evaluations, SPSA performance improves (as expected) and converges to schedules of the same quality as BO. Because, SPSA uses a proxy for the gradient to suggest the next parameters, we expect SPSA to achieve higher accuracy in fine-tuning the parameters, ultimately making it the preferred method in cases where many evaluations are possible. In addition, in Appendix~\ref{app:independent-controls} we compare the efficiency of BO with results from Hedge \textit{et al.}~\cite{hegde_genetic_2022}, based on a genetic algorithm to optimize annealing schedules for QA in the $p$-spin model. As shown in Fig.~\ref{fig:hedge-comp} we find that BO is able to find schedules outperforming those found by the genetic algorithm, using an order of magnitude fewer iterations.

\begin{figure}[htb]
\subfloat[]{%
    \hspace{-0.6cm}
  \includegraphics{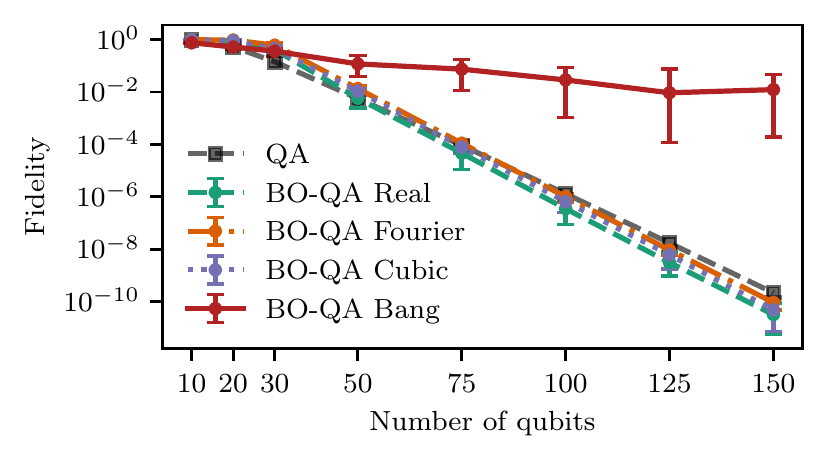}%
  \label{fig:fidelity-from-energy}
}
\vspace{-0.8cm}
\subfloat[]{%
\hspace{-0.6cm}
  \includegraphics{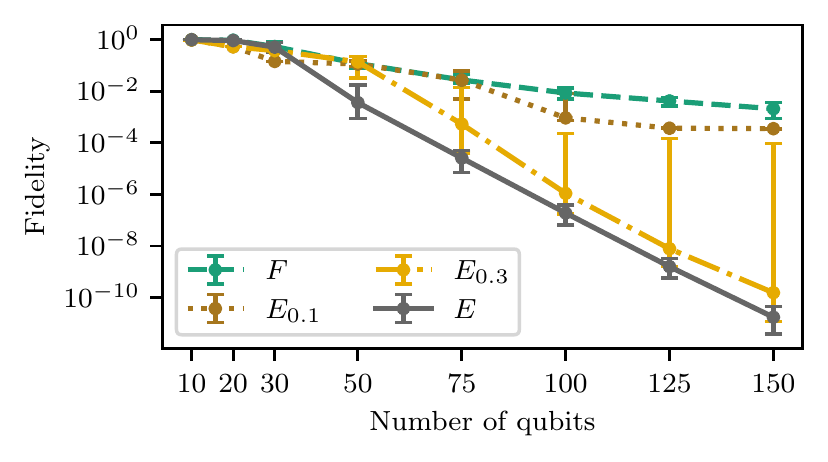}%
  \label{fig:fidelity-from-foms}
}
\caption{(a) Scaling of the fidelity, when energy is used as the figure of merit. Only the Bang-bang schedule is able to improve on the linear schedule.
(b) Scaling of the ground-state fidelity when different figures of merit are used. Here $E_x$ corresponds to the energy of the top $x$-th quantile (in terms of energy) of the measured bit strings, $E$ to the energy, and $F$ to the fidelity.}
\label{fig:figures-of-merit}
\end{figure}

We next investigate the mechanisms underlying the improved schedules found by BO. Because we use a dependent parametrization of the schedules (see Eq.~\eqref{eq:pspin-qa}), the minimal instantaneous gap that these parametrizations encounter during the protocol is the same as in the linear schedule; see Fig.~\ref{fig:qa-gap-paths}b. The improvement must therefore stem either from the ability of BO to learn \emph{adiabatic} schedules that adapt their rate of change to the spectral gap profile, or alternatively, by finding \emph{diabatic} paths, forgoing the requirement of adiabaticity. We find in our simulations that the latter is the predominant case for the $p$-spin model; see a typical example in Figure~\ref{fig:qa-gap-paths}. Here, the path found is in fact highly non-adiabatic, as the schedule traverses the region with a minimal spectral gap multiple times due to its nonmonotonicity. This results in diabatic transitions in and out of the instantaneous ground state, as seen in panel (c) of Fig.~\ref{fig:qa-gap-paths}. However, BO is able to exploit these diabatic transitions such that the obtained fidelity is approximately $35\times$ larger than the one obtained by the linear path.

Finally, we analyze the importance of selecting an appropriate figure of merit (FoM) for BO. In a realistic setting (e.g., when solving a combinatorial optimization problem), using the ground-state fidelity as the FoM is not possible because the ground state of the target Hamiltonian is typically not known a priori. Therefore, we also analyze the performance when proxies for the fidelity, such as the energy, are used as the FoM for the optimizer. In Fig.~\ref{fig:fidelity-from-energy}, we show results when the energy serves as the feedback signal seen by the optimizer. Strikingly, for all parametrizations except for the optimal bang-bang parametrization, the optimizer is unable to learn schedules that would improve on the linear schedule in terms of fidelity. This finding led us to consider different FoMs, which we constructed by taking only the top $x$-th quantile of the measured bit strings in terms of energy -- we denote this by $E_x$. Results from optimizing the real parametrization in terms of different FoMs are shown in Fig.~\ref{fig:fidelity-from-foms}. We find that taking the top 30\% or 10\% of the measured bit strings improves the obtained fidelities dramatically. Intuitively, it is reasonable to expect that biasing the optimizer to increase the probability of measuring very low energy states as opposed to optimizing for the \emph{global} energy expectation value for the observed state is a better proxy for the fidelity. Similar observations about the importance of choosing a good FoM in variational quantum algorithms were made in Ref.~\cite{barkoutsos_improving_2020}, where Conditional Value at Risk was proposed as a possible alternative.

\subsubsection{Reverse Annealing}

\begin{figure*}[ht]
\centering
\subfloat[]{%
\includegraphics{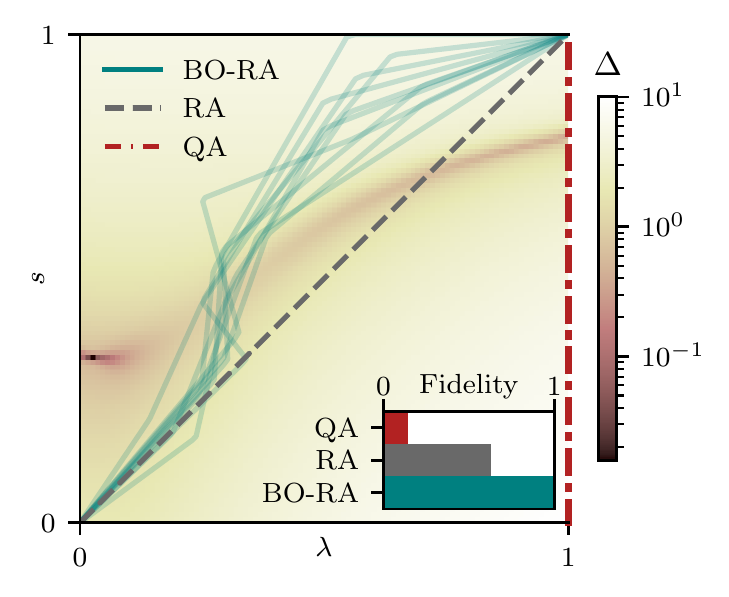}%
\label{fig:ra-paths}%
}
\qquad
\subfloat[]{%
\includegraphics{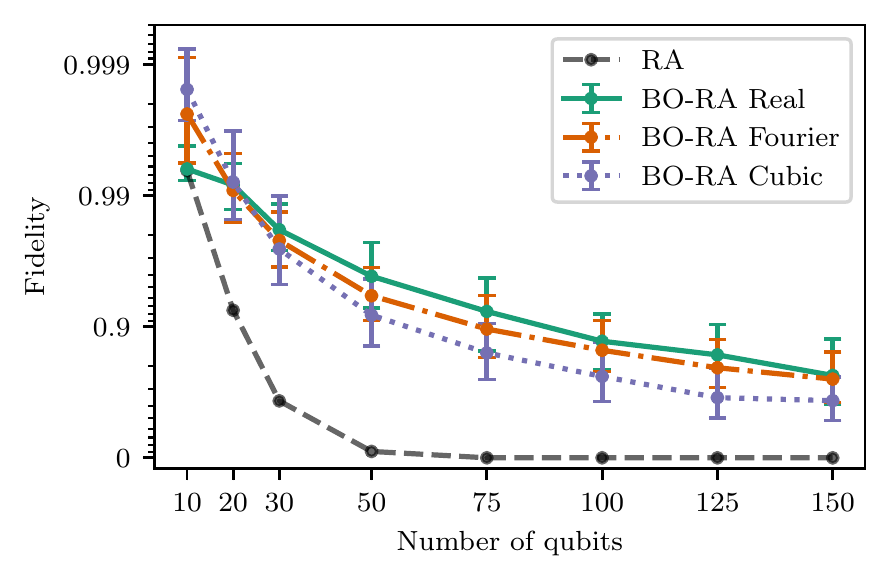}%
\label{fig:ra-scaling}%
}
\caption{(a) The BO-RA, RA and QA ($\lambda=1$) paths drawn on top of the spectral gap $\Delta$ landscape, for a system with $30$ spins. A real-space parametrization is used with $6$ parameters specifying the three intermediate points. BO successfully finds a path through the region where the gap is larger, leading to a larger fidelity (see inset). The median fidelity of the $10$ runs shown in the diagram (light teal) is $0.996$. 
(b) Scaling of the ground state fidelity for different parametrizations. BO is able to find schedules with good fidelities for large system sizes, where standard RA fails (see black dashed line). 
}
\end{figure*}

We next turn our attention to Reverse Annealing (RA) as applied to the $p$-spin model. Here we limit ourselves to the case of $c=0.8$, meaning that in the initial state $80\%$ of the spins are aligned with the final ground state. While RA has been shown to enable an exponential speedup compared to QA, careful fine-tuning of the parameters is required to fully unlock this speedup, if only linear schedules $\lambda(t)=s(t)=t/\tfinal$ are considered~\cite{yamashiro_dynamics_2019}. To see this, consider the phase diagram in Fig.~\ref{fig:ra-paths}. Here, the linear path from $\lambda=s=0$ to $\lambda=s=1$ fails to leverage the part of the phase diagram where the gap is larger. In contrast, the BO-designed schedules shown in Fig.~\ref{fig:ra-paths} consistently find paths going through the opening in the gap landscape. This \textit{opening} corresponds to the part of the phase diagram where the first-order phase transition is avoided in the thermodynamic limit -- in the case of finite system sizes, this corresponds to the exponential speedup demonstrated in Ref.~\cite{yamashiro_dynamics_2019}, which we are able to harness using BO. This results in improved fidelities over QA (corresponding to a path with $\lambda(t)=1$) and non-optimized RA. 

The importance of finding a path with favorable spectral properties is apparent in the system size scaling results presented in Fig.~\ref{fig:ra-scaling}. At the largest system size considered here (with $N=150$ spins), the median fidelity of the real space BO runs is found to be $\sim 0.76$, compared to $3.7\cdot 10^{-6}$ in the case of a linear RA schedule, yielding an improvement of more than $5$ orders of magnitude.

\subsubsection{Effects of noise}
\begin{figure}[htb]
    \includegraphics{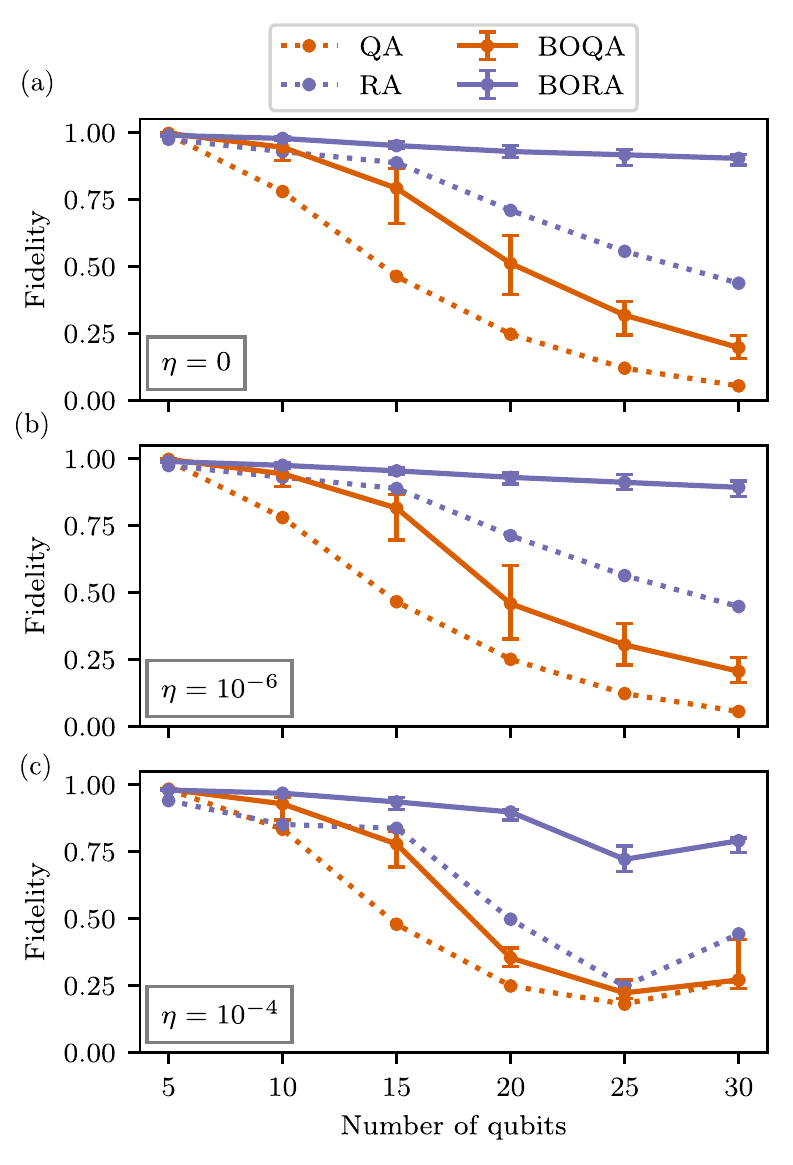}
    \caption{Results on RA for the $p$-spin model with $30$ spins and collective dephasing noise, the strength of which is captured in the magnitude of the $\eta$ parameter. $\eta=0$ correspond to unitary evolution. A total annealing time of $\tfinal=20$ is used.}
    \label{fig:pspin-decoherence}
\end{figure}

We now analyze the effects of noise. For context, it has previously been shown by Passarelli \textit{et al.}~\cite{passarelli_standard_2022} that RA loses some of its computational edge over QA in the presence of noise. In particular, in Ref.~\cite{passarelli_standard_2022} the effects of dephasing noise were analyzed. The authors showed that QA appears to be more robust against both collective and independent dephasing, with QA outperforming RA in terms of time-to-solution in the presence of these noise channels.

To this end we set up simulations of the adiabatic master equation given in Eq.~\eqref{eq:ame}, and analyze the ability of BO to find good annealing schedules in the presence of collective dephasing of various strengths. Figure~\ref{fig:pspin-decoherence} displays the results for several values of the noise parameter $\eta$ controlling the strength of the coupling to the environment (larger $\eta$ corresponds to noisier systems). Here we use a total annealing time of $\tfinal=20$, and a real parametrization with a total of $4$ parameters, for both RA and QA. In line with the results of Ref.~\cite{passarelli_standard_2022}, we find that with increasing $\eta$, unoptimized RA fidelities approach those obtained by QA (see dotted lines). Additionally, we see that QA may even benefit from the presence of noise: notice how fidelities at the largest system size (with $N=30$ spins) increase with increasing $\eta$. This effect has previously been documented in Ref.~\cite{passarelli_reverse_2020}. Most importantly for the present study, we find that BO is able to find paths that overcome this limitation of RA, thus further motivating the adoption of BO in experimental real-world scenarios typically prone to noise.

Interestingly, the improvement of BO for QA schedules is diminishing for larger values of $\eta$. This could be due to the limited flexibility of the dependent parametrization used in the simulations or the qualitatively different spectral properties of the QA and RA Hamiltonians. To investigate the underlying mechanisms, we perform numerical simulations of the same system, with independent controls for QA. Results, summarized in Figure~\ref{fig:dephasing-independent}, indicate that the additional flexibility offered by an independent parametrization indeed enables BO to design better schedules in the presence of noise.

\subsection{Maximum Independent Set}
\label{sec:results-mis}

\begin{figure}[htb]
\subfloat[]{%
  \includegraphics{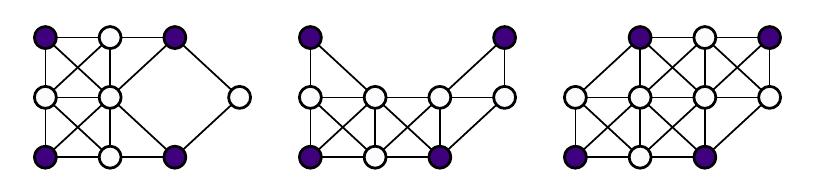}%
  \label{fig:smallgraphs-schematic}
}
\vspace{0.2cm}
\subfloat[]{%
  \includegraphics{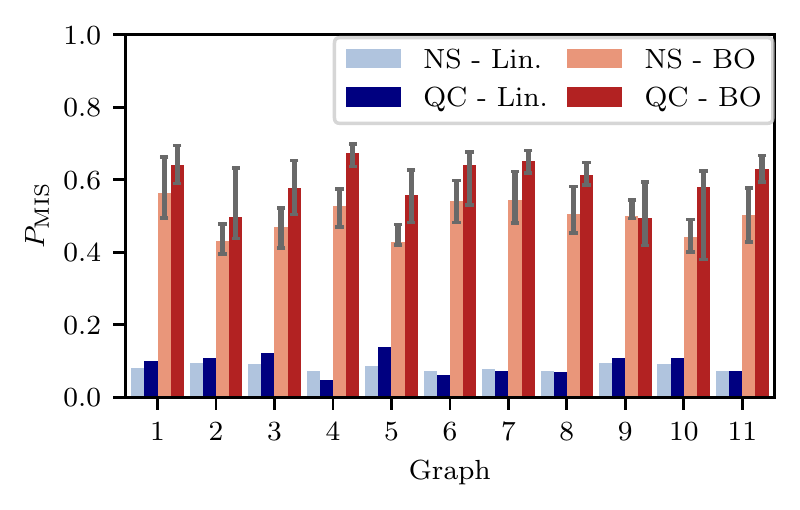}%
  \label{fig:smallgraphs-barplot}
}
\caption{(a) Example of small toy graphs, with $9$ -- $10$ nodes. Nodes colored in purple represent the (unique) solution of the MIS problem.
(b) Success probabilities $P_{\mathrm{MIS}}$ of solving the MIS problem using a neutral atom quantum device, for both a linear schedule (blue hues) and BO schedules (red hues). Lighter and darker shades correspond to results obtained in numerical simulations (NS) and on actual quantum hardware (QC), respectively.}
\label{fig:smallgraphs}
\end{figure}

Next, we turn our attention to solving the MIS problem on Rydberg atom arrays, as available online through Amazon Braket via the QuEra device. We first analyze numerical simulations of the experimental platform and then compare these to actual runs on the QuEra Aquila device. Finally, we extend our analysis to graph sizes for which classically simulating the quantum many-body evolution as described by the Hamiltonian in Eq.~\eqref{eq:rydberg-hamiltonian} becomes infeasible.

We first direct our attention to small graphs, to gain an intuitive insight into efficient schedule Ansaetze. Specifically, we generate all $11$ non-isomorphic instances of UDGs with radius $\sqrt{2}a<R_{\mathrm{d}}<2a$, where $a$ is the lattice constant of the underlying $4\times 3$ square lattice on which 9 or 10 nodes are placed (leaving between $2$ and $3$ vacancies), such that the instances have a unique ground state (verified via a brute-force search of all possible configurations). We show three randomly-selected graphs in Fig.~\ref{fig:smallgraphs-schematic} and  choose the lattice constant to be $a=\SI{5.3}{\micro\metre}$, which we have found to work best with the considered quantum hardware.

Aiming for low-complexity Ansaetze with a modest number of parameters, it is vital to utilize the corresponding parameters efficiently. In particular, this holds for Rydberg atoms, where in principle (at minimum) two schedules (for global detuning and Rabi drive) can be tuned largely independently. Therefore, we first consider a relatively simple parametrization that uses the initial and final values ($\deltainitial$ and $\deltafinal$, respectively) of the detuning schedules, and the ramp-up time $\tau_{\Omega}$ as well as the maximal Rabi frequency $\Omega_{\max}$ specifying $\Omega(t)$ -- see Figure~\ref{fig:rydberg-schedules}. In all experiments on the toy graphs the total duration of the protocol was set to $\tfinal=\SI{0.7}{\micro\second}$. Because both the $\Delta$ and the $\Omega$ schedules consist of exclusively linear line segments, we refer to this Ansatz as the \emph{BO-Linear} parametrization. The initial parameters for this linear parametrizations are partially inspired by Ref.~\cite{ebadi_quantum_2022} and partially chosen such that the Rydberg blockade radius $R_{\mathrm{b}}$ (see Sec.~\ref{sec:mis-rydberg}) is in the desired range $\sqrt{2}a<R_{\mathrm{b}}<2a$; see Appendix~\ref{app:mis} for details. We optimize the linear parametrization ten different times for each of the 11 small graphs using classical simulations and then evaluate the schedules on quantum hardware for comparison. As the figure of merit optimized by BO, we use the energy of the top $50^{\mathrm{th}}$ percentile (dubbed $H_{0.5}$). Finally, we emphasize that in this work we focus on the performance of the quantum device and have thus refrained from using classical postprocessing as done in Ref.~\cite{ebadi_quantum_2022}.

We compare the different schedules in terms of the probability $P_{\mathrm{MIS}}$ of measuring the solution to the MIS problem. Results shown in Fig.~\ref{fig:smallgraphs-lineplot} show that significant improvements can be made over the naive choice of parameters for the linear schedule (blue-shaded lines, average $P_{\mathrm{MIS}}\sim 0.08$) by just optimizing its parameters (green-shaded lines, average $P_{\mathrm{MIS}}\sim 0.35$). Additionally, we show in Fig.~\ref{fig:gridgraphs-colorplot} how the improvement in the schedules manifests itself in that the expectation value of the Rydberg excitations $\langle\hat{n}_i\rangle$ of the $i$-th node matches the MIS (shown in Fig.~\ref{fig:smallgraphs-schematic}) ever closer. Finally, the numerical simulation results are found to be in good agreement with the results from the QuEra hardware.  Moreover, we observe that the optimal parameters of the $\Omega(t)$ schedule concentrate significantly at the maximal allowed value by the hardware for $\Omega_{\max}$ and for shorter values of $\tauomega$ (see Fig.~\ref{fig:mis-parameters-hist} in Appendix~\ref{app:mis}). The latter value may intuitively be understood as shorter values of $\tauomega$ correspond to extended time intervals $\tfinal - 2\tauomega$ during which $\Delta$ is varied (see Fig.~\ref{fig:rydberg-schedules}). Therefore, the system has a higher chance of adhering to the (quasi-)adiabatic limit, because the rate of change of $\Delta$ is smaller.

Motivated by these findings, we consider yet another parametrization, in which only the $\Delta(t)$ schedule is varied, and the $\Omega(t)$ parameters are fixed to $\Omega_{\max}=\SI{15.8}{\mega\hertz}$ and $\tau_{\Omega}=0.1\tfinal$. We parametrize the $\Delta(t)$ schedule using a real-space parametrization with two intermediate points as described in Sec.~\ref{sec:schedule-parametrizations} interpolating between $\deltainitial$ and $\deltafinal$, as shown in Fig.~\ref{fig:rydberg-schedules}. We further pass the obtained schedule through a low-pass filter, as previously suggested in Ref.~\cite{ebadi_quantum_2022}. The results obtained using this parametrization are shown in Fig.~\ref{fig:smallgraphs-barplot}, indicating that improvements of an order of magnitude in the probability of measuring the MIS state $P_{\mathrm{MIS}}$ are possible using this approach. Furthermore, we again observe good agreement between numerical simulations of the optimized schedules and the corresponding results obtained on the real hardware.

\begin{figure*}[ht]
\centering
\includegraphics[width=\textwidth]{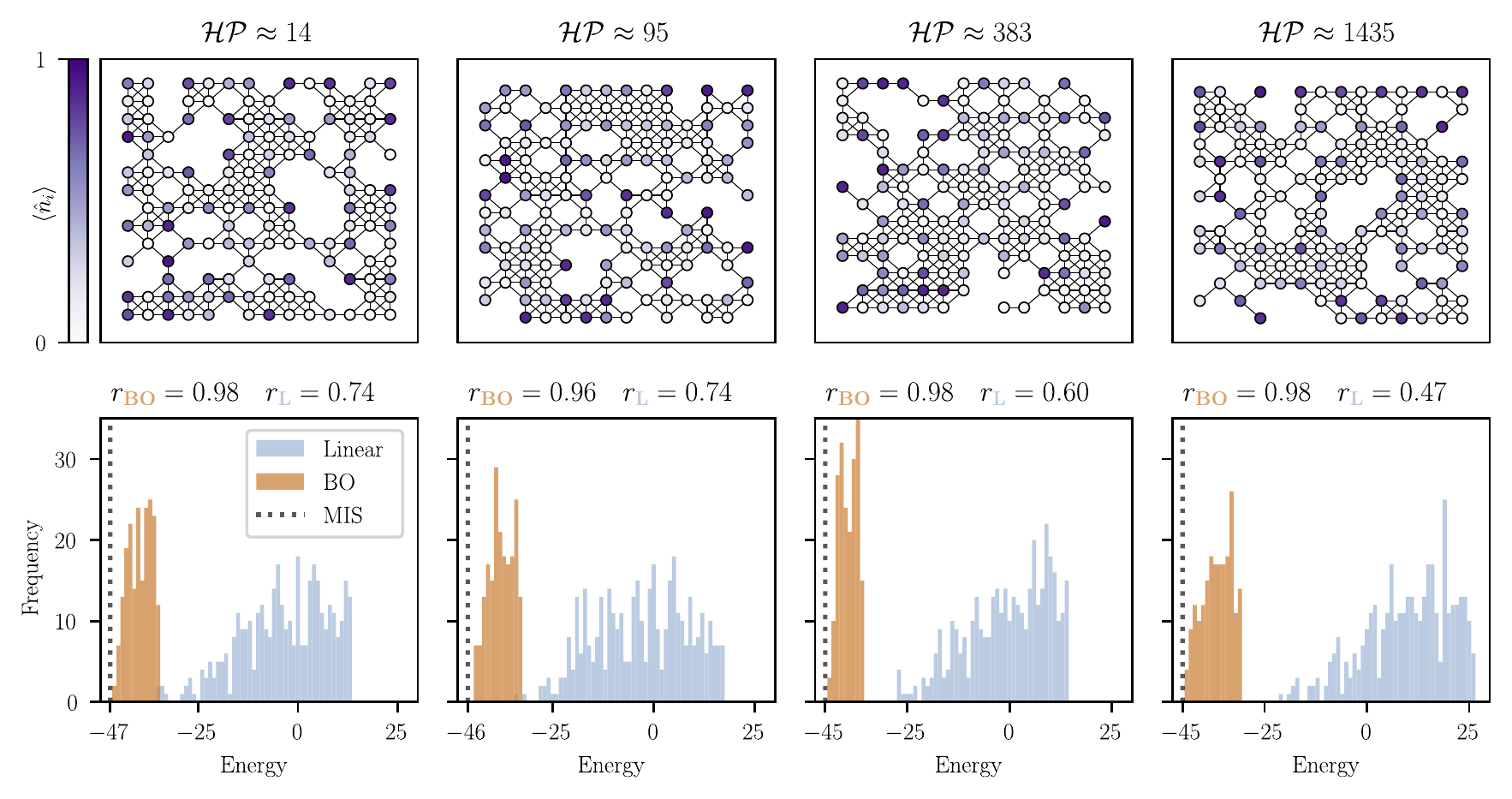}
\caption{
Example data from the QuEra Aquila device. Above: Unit-disk graph instances with 137 nodes partially filling underlying square lattices with a lattice constant of \SI{5.3}{\micro\metre}, and nearest and next-nearest coupling. The instances are sorted according to their hardness parameter $\hp$ (from left to right). The node's color indicates the probability $\langle\hat{n}_i\rangle$ of the $i$-th node being assigned to the independent set ($\langle\hat{n}_i\rangle=1$ when the node is purple), as determined by the BO protocol. Several nodes appear (approximately) equally likely in and out of the independent set. This is due to the degeneracy of the low-energy solutions which require different nodes to be part of the independent set. 
Bottom: Histograms of the top 50\% of the obtained energies (as defined in Eq.~\eqref{eq:mis-cf}) with an unoptimized linear schedule (blue), and the BO schedule (orange). The optimal solution of size $|\mathrm{MIS}|$ is indicated by the vertical dotted lines. $r_{\text{BO}}$ and $r_{\text{L}}$ correspond to the approximation ratios of the linear and BO schedules, respectively. There is a clear improvement on the data when BO is used.}
\label{fig:hardgraphs}
\end{figure*}

We now test the pipeline described above on larger instances. These are generated in a similar fashion to the small graphs in Fig.~\ref{fig:smallgraphs-schematic}, in that an underlying square lattice is generated and subsequently filled with nodes, connecting only neighboring and diagonal nodes (i.e., nearest and next-nearest neighbors). Specifically, we consider a grid of size $14\times 14$ with a lattice constant of $a=\SI{5.3}{\micro\metre}$, positioning nodes on a randomly chosen 137 sites, corresponding to a filling of $\sim 70\%$. It was previously shown that the performance of simulated annealing~\cite{Kirkpatrick1983}, a commonly used classical technique for combinatorial optimization, is limited by the ratio of the degeneracies of nearly optimal and optimal configurations~\cite{ebadi_quantum_2022}. This is captured by the so-called hardness parameter 
\begin{equation}
\mathcal{HP}:=\frac{N_{|\mathrm{MIS}|-1}}{|\mathrm{MIS}| \cdot N_{|\mathrm{MIS}|}},    
\end{equation}
where $N_{M}$ is the degeneracy of configurations corresponding to independent sets of size $M\leq |\mathrm{MIS}|$. Using the tropical tensor network algorithm described in Refs.~\cite{liu_computing_2022, Liu_2021} we can compute the hardness parameter $\hp$, and quantify the difficulty on a per instance basis. In Fig.~\ref{fig:hardgraphs} a collection of graphs with 137 nodes is shown, which have dramatically different hardness parameters, despite appearing qualitatively similar to the human eye. We run the BO pipeline to optimize the $\Delta(t)$ schedule for these graphs and use $100$ shots to obtain an estimate of $H_{0.5}$. Here, we use $\tfinal=\SI{4.0}{\micro\second}$, which is the maximal duration implementable on the QuEra Aquila device. While we are not able to solve the instances to optimality, the obtained solutions (taken from $800$ samples) are significantly improved compared to the linear schedule; see bottom row in Fig.~\ref{fig:hardgraphs}. We quantify this by reporting the approximation ratios 
\begin{equation}
\label{eq:approx-ratio}
r:= \frac{|\min_{\xbf\in M} H(\xbf)|}
{
|\mathrm{MIS}|
},
\end{equation}
where $M$ is the set of bit strings obtained from the device, and $H(\xbf)$ is the MIS cost function, as defined in Eq.~\eqref{eq:mis-cf}. For each graph, $r_{\text{BO}}$ and $r_{\text{L}}$ denote the approximation ratios of the best bit string obtained by the BO and Linear protocols, respectively. It should be noted that the approximation ratio $r$ includes the independence constraint as a \emph{soft} constraint with a penalty $\alpha$ -- as such, the energies reported in Figure~\ref{fig:hardgraphs} can in principle correspond to non-independent sets. However, for BO-designed protocols, the best sets used to compute $r_{\text{BO}}$ are always independent, while for the linear schedule the obtained bit strings correspond either to very small independent sets, or exclusively to solutions that violate the independence constraint. This further highlights the dramatic impact of BO on the quality of the obtained solutions.

While it is difficult to draw conclusions from a comparably small data set, our results indicate that increasing difficulty (in terms of $\hp$) has a more detrimental effect on the quality of bit strings obtained in the linear protocol than in the BO protocol. This is in line with previous studies~\cite{ebadi_quantum_2022}, which found substantial correlation between $\hp$ and the minimum spectral gap of the Rydberg Hamiltonian in Eq.~\eqref{eq:rydberg-hamiltonian}. It appears as if the BO-designed schedules can (partially) overcome this by adjusting the speed of the evolution to the spectral gap profile or by leveraging diabatic transitions as in the case of the $p$-spin model.

Although BO is specifically designed for resource-efficient optimization, we find at the time of writing this manuscript that deploying the algorithm on real hardware is difficult due to the relatively long readout times~\cite{Henriet_2020}. To reproduce the procedure devised in Ref.~\cite{ebadi_quantum_2022} for \emph{a single graph} would require several weeks, because their pipeline requires approx.~$10^5$ measurements from the quantum device to optimize the schedules. This simultaneously gives merit to our BO pipeline, which uses an order of magnitude less resources and highlights the need for a more holistic evaluation of quantum hardware, which includes wall-clock running times~\cite{finzgar_quark_2022}.

\section{Discussion}
\label{sec:discussion}

In this work we demonstrated how BO can be used to efficiently design quantum annealing schedules in a diverse set of scenarios, using a relatively low number of evaluations (queries of the cost function). While BO is able to learn suitable parameters for a variety of schedules, even in the presence of noise, stark differences in performance between different parametrizations indicate that the particular choice of a schedule Ansatz is important and depends on the problem (class) in consideration. Similarly, the choice of the figure of merit passed to the optimizer strongly influence the quality of the results. These design choices will depend not only on the specific problem class but also on the resources available (e.g., time requirements). 

For the $p$-spin model BO-designed QA schedules are able to learn diabatic mechanisms, which dramatically improve the success probability. However, the degree to which this can be generalized to other systems is unclear, especially because the model is highly symmetric and has been shown to be amenable to diabatic ground state preparation in the past~\cite{wauters_polynomial_2020, garcia-pintos_lower_2022}. Here, we provide numerical evidence that BO-designed RA schedules can overcome the previously identified shortcomings of RA in the presence of noise~\cite{passarelli_standard_2022}. This potentially motivates a new research direction of optimized RA schedules. It would likewise be interesting to identify the underlying reasons for the inability of BO to significantly improve QA schedules when noise is present. This may indicate that the diabatic mechanisms are not robust against dephasing noise or that noise somehow hampers the ability of BO to find them. Nonetheless, results shown in Fig.~\ref{fig:dephasing-independent} suggest that it is possible to enhance the robustness of QA schedules by considering a more flexible parametrization with independent controls, as demonstrated in Appendix~\ref{app:independent-controls}. While these results might be due to the specifics of the $p$-spin model, they nonetheless offer an interesting probe into the mechanisms underlying the noise-robustness of (a)diabatic paths.

Furthermore, we have shown how to apply our BO pipeline for solving the MIS problem using neutral atom quantum computers. Specifically, we were able to achieve substantial improvements over the naive linear adiabatic protocol, even for relatively complex graphs with more than 100 nodes. We also find that schedules designed in numerical simulations possess a high degree of transferability to real quantum hardware, where they exhibit comparable performance. This is in line with the relatively low error rates of this qubit modality~\cite{Henriet_2020}, especially in the analog mode of operation utilized here. Our results along with other recent experimental findings~\cite{ebadi_quantum_2022} highlight the promise of variational quantum adiabatic algorithms, such as those considered in this manuscript, as a direction for future research in quantum optimization. In particular, our BO pipeline has demonstrated its ability to optimize quantum protocols resource-efficiently, making it a valuable practical tool for exploring variational quantum algorithms.
 
\section{Outlook}
\label{sec:outlook}

Finally, we highlight possible extensions of research going beyond our present work. The BO pipeline proposed here can be amended in various ways, e.g., by using different choices of kernel and acquisition functions, considering different FoMs (e.g., conditional value at risk~\cite{barkoutsos_improving_2020}), or different parametrizations. Moreover, one could envision schemes combining BO with other optimizers with complementary properties. One such possibility would be to combine it with an optimizer better suited to fine-tune the parameters, which would derive its initialization from the model of the FoM constructed by BO.

As BO has proven to be capable of efficiently optimizing QA schedules, it is natural to consider how it may aid the design of protocols based on extensions of QA, such as including non-stoquastic~\cite{Nishimori_2017} or inhomogeneous transverse field~\cite{Susa_2018} driver Hamiltonians. Additionally, extending the analysis here beyond the $p$-spin model would be worthwhile, in particular to determine whether its ability to find diabatic and RA paths can be transferred to other systems. In addition to numerical simulations, experiments on quantum annealers based on superconducting flux qubits could be performed~\cite{Johnson_2011,King_2022}. Over in the realm of classical optimization algorithms, one might even consider designing the temperature schedules for annealing and tempering algorithms using BO.

Finally, a systematic analysis of BO-designed schedules for solving the MIS problem on UDGs in terms of the hardness parameter is deferred to future work. Our inability to conduct such systematic studies could be remedied by devising yet more efficient protocols to optimize the schedules, e.g., by tailoring the schedule parametrization to the typical spectral gap profiles of the Rydberg Hamiltonian or by an increased capacity to obtain measurement results from neutral atom devices. Rydberg-based special purpose quantum devices are at their infancy and, as such, there is still a lot to be learned.

\section*{Acknowledgements} 
\label{sec:acknowledgements}
J.R.F.~acknowledges helpful discussions with Gianluca Passarelli on setting up numerical simulations of the AME.
The authors thank Gili Rosenberg for carefully reviewing the manuscript.
H.G.K.~would like to thank H.~B.~H.~Urweisse for launching this collaboration.
The work of H. N. is based on a project JPNP16007 commissioned by the New Energy and Industrial Technology Development Organization (NEDO), Japan.

\bibliographystyle{apsrev4-1}
\bibliography{lit, references}

\appendix

\section{Independent controls} 
\label{app:independent-controls}

\begin{figure}[htb]
\includegraphics{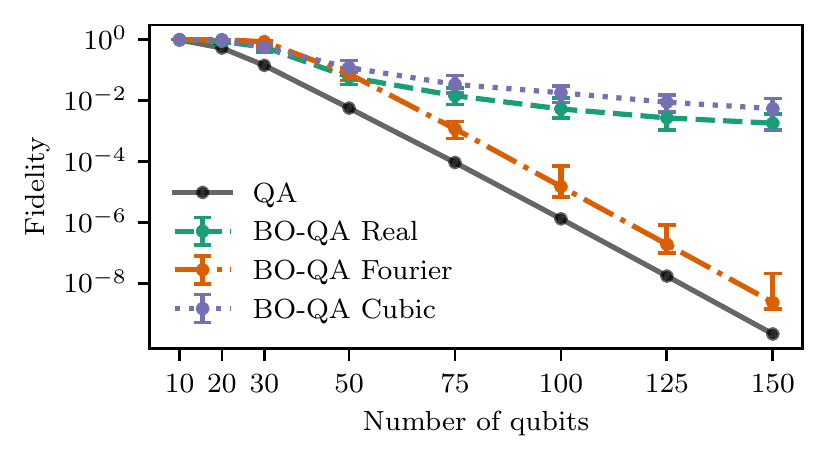}
\caption{Fidelity of BO schedules, when independent control functions are used. Otherwise, parameters are kept the same as in the results presented in Fig.~\ref{fig:fig1}.}
\label{fig:qai-overlaps}
\end{figure}

In the case of independent controls, Eq.~\eqref{eq:pspin-qa} is generalized as
\begin{equation}
\label{eq:pspin-qa-independent}
    \hhat(t)
    =
    u_1(t)\hhat_{\mathrm{TF}}
    +
    u_2(t)\htarget,
\end{equation}
where $u_1(0)=u_2(\tfinal)=0$ and $u_1(\tfinal)=u_2(0)=1$. Such a parametrization requires more parameters, as no parameters are shared between $u_1(t)$ and $u_2(t)$. However, it might be beneficial to use such parametrizations when more resources are available, because it enables a higher flexibility for schedule design. For example, in contrast to the dependent parametrization in Eq.~\eqref{eq:pspin-qa}, different choices of independent schedules might encounter different minimal spectral gaps along the way. In Fig.~\ref{fig:qai-overlaps} we show the same experiments described in Fig.~\ref{fig:qad-scaling}, however, with the 4 parameters distributed evenly between $u_{1,2}$.

\begin{figure}[htb]
\includegraphics{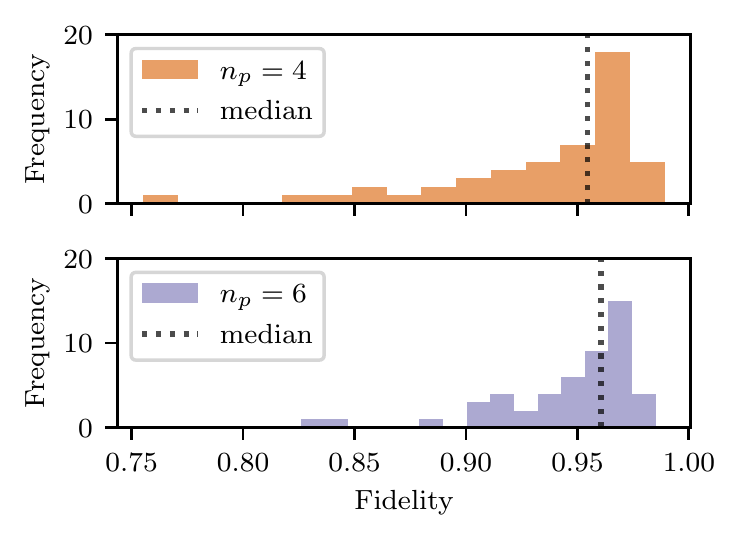}
\caption{Histogram of the ground-state fidelities for $50$ runs of BO with independent controls. Using the same parameters, Hedge \textit{et al.}~\cite{hegde_genetic_2022} report a median fidelity of approximately $0.90$, which is surpassed by a cubic parametrization with both four and six parameters. Additionally, BO requires approximately an order of magnitude fewer evaluations of the optimization objective.}
\label{fig:hedge-comp}
\end{figure}

Such a parametrization was likewise used in Ref.~\cite{hegde_genetic_2022}, where a genetic algorithm was used to optimize annealing schedules for the $p$-spin model. Here, we compare the results they obtained for the $p$-spin model with $p=3$, a system size of $15$ spins with $\Gamma=1$, and using fidelity as the FoM, which was also the case in Ref.~\cite{hegde_genetic_2022}.
Using a modified BO pipeline from Sec.~\ref{sec:bo-pipeline} with $80$ random parameter evaluations, and $420$ iterations, we outperform the genetic algorithm used in Ref.~\cite{hegde_genetic_2022} using a total of $80 + 420 = 500$ evaluations, which is an order of magnitude less in resource requirements -- see Figure~\ref{fig:hedge-comp}. We use the cubic parametrization with $n\in\left\{4, 6\right\}$ parameters and $\zeta=2(n + 1)$.

\begin{figure}[htb]
\includegraphics{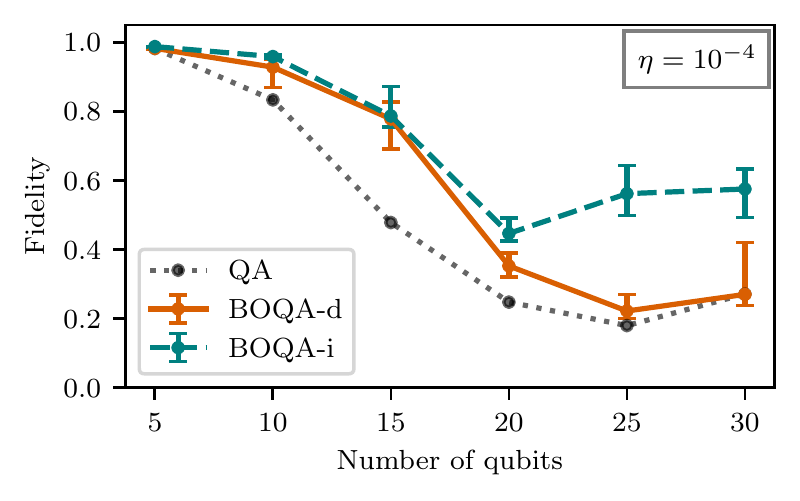}
\caption{Results of with collective dephasing for dependentent (-d) and independent (-i) controls, for the case of $\eta=10^{-4}$. Other parameters are kept as in Figure~\ref{fig:pspin-decoherence}.}
\label{fig:dephasing-independent}
\end{figure}

Finally, to identify the mechanisms underlying the shortcomings of the dependent parametrization for stronger collective dephasing---as seen in the case of $\eta=10^{-4}$ in Fig.~\ref{fig:pspin-decoherence}---we show here results of numerical simulations for independent controls. In Fig.~\ref{fig:dephasing-independent}, the obtained fidelities are shown and compared to the dependent case. We only show results for $\eta = 10^{-4}$ because dependent and independent controls exhibit similar behaviour for lower noise levels. The results in Fig.~\ref{fig:dephasing-independent} hint towards the fact that the additional flexibility of the independent parametrization is indeed beneficial. However, the systematic analysis of such effects that would be required to underpin the explanation for these observations is outside the scope of this paper.

\section{Adiabatic Master Equation}
\label{app:ame}

In this Appendix we provide additional details about our implementation of the Adiabatic Master Equation (AME) supplementing Section~\ref{sec:pspin-noise}. We closely follow the approach outlined by Passarelli \textit{et al.}~\cite{passarelli_standard_2022}, which we briefly review here.
For further details, we refer the reader to the original publication. Restating the AME [Eq.~\eqref{eq:ame}]
$$
    \partial_t \hat{\rho}(t)
    =
    -i\left[\hhat(t)+\hhat_{\mathrm{LS}}(t), \hat{\rho}(t)\right]
    +
    \mathcal{D}[\hat{\rho}(t)],
$$
we next specify the exact form of the terms. This form of the AME arises when one considers a system-bath coupling of the form $\hhat_{\mathrm{SB}} = \hat{A}\otimes\hat{B}$, where $\hat{A}$ and $\hat{B}$ are operators acting on the Hilbert space of the principal system $\mathcal{H}_S$ and the bath $\mathcal{H}_B$, respectively. We limit ourselves to considering only collective dephasing, corresponding to $\hat{A}=\spin{z}$. Because $\hat{A}$ is permutationally invariant, this allows us to work in the reduced Hilbert space of maximal $z$ angular momentum, as discussed in Sec.~\ref{sec:pspin-model}. The time-dependent Lindblad operators arising from this coupling can then be obtained as:
\begin{equation}
\label{eq:lindblad-op}
    \lhat_{ab}(t)
    =
    \ket{E_a(t)}\bra{E_a(t)}
    \hat{A}
    \ket{E_b(t)}\bra{E_b(t)},
\end{equation}
where $\ket{E_a(t)}$ are the instantaneous eigenstates of $\hhat(t)$ corresponding to the eigenenergies $E_a(t)$. Using Eq.~\eqref{eq:lindblad-op} we can finally state the Lamb-Shift Hamiltonian:
\begin{equation}
\begin{aligned} 
H_{\mathrm{LS}}(t) 
= & 
\sum_{a, b, a \neq b}
S\left(\omega_{b a}(t)\right)
\lhat_{a b}^{\dagger}(t) \lhat_{a b}(t) 
\\ 
& +
\sum_{a, b}
S(0)
\lhat_{a a}^{\dagger}(t) \lhat_{b b}(t),
\end{aligned}
\end{equation}
and the dissipator:
\begin{equation}
\begin{aligned}
\mathcal{D}[\hat{\rho}(t)]
= &
\sum_{a, b, a \neq b}
\gamma\left(\omega_{b a}(t)\right)
\left(
\lhat_{a b}(t) \hat{\rho}(t) \lhat_{a b}^{\dagger}(t)
\right. 
\\ &
\left.-
\frac{1}{2}\left\{
\lhat_{a b}^{\dagger}(t) \lhat_{a b}(t), \hat{\rho}(t)
\right\}
\right) 
+
\\ &
\sum_{a, b}
\gamma(0)
\left(
\lhat_{a a}(t) \hat{\rho}(t) \lhat_{b b}^{\dagger}(t)
\right. 
\\ &
\left.-\frac{1}{2}
\left\{
\lhat_{a a}^{\dagger}(t) \lhat_{b b}(t), \hat{\rho}(t)
\right\}
\right).
\end{aligned}
\end{equation}
Here we have used $\omega_{ab}(t) = E_a(t) - E_b(t)$ and the anticommutator $\{\hat{X}, \hat{Y}\}= \hat{X}\hat{Y} + \hat{Y}\hat{X}$. Additionally, $\gamma(\omega)$ is the relaxation rate, and $S(\omega)$ is its Hilbert transform. As done in Ref.~\cite{passarelli_standard_2022}, we consider an Ohmic bath of the form
\begin{equation}
    \gamma(\omega)
    =
    2\pi\eta\frac{
    \omega\exp(-|\omega|/\omega_{\mathrm{c}})
    }{
    1-\exp(-\beta\omega)
    }.
\end{equation}
Setting $k_{\mathrm{B}}=1$ we use $\beta = 1/T$ with $T=\SI{12}{\milli\kelvin}$ and $\omega_{\mathrm{c}}=8\pi$. We simulate the AME in the density matrix picture using the \texttt{HOQST} library~\cite{chen_hamiltonian_2022}. Because we consider system parameters that are still relatively close to the adiabatic limit, we use an iterative method~\cite{calvetti1994implicitly} and only keep the lowest $30$ instantaneous eigenstates to form the operators in Eq.~\eqref{eq:lindblad-op}, as we expect that higher excitations will not significantly influence the outcome. We verified that the results do not change when more levels (e.g., 60) are considered.

\section{MIS on neutral atom quantum computer: Implementation details}
\label{app:mis}

\begin{figure}[htb]
\includegraphics{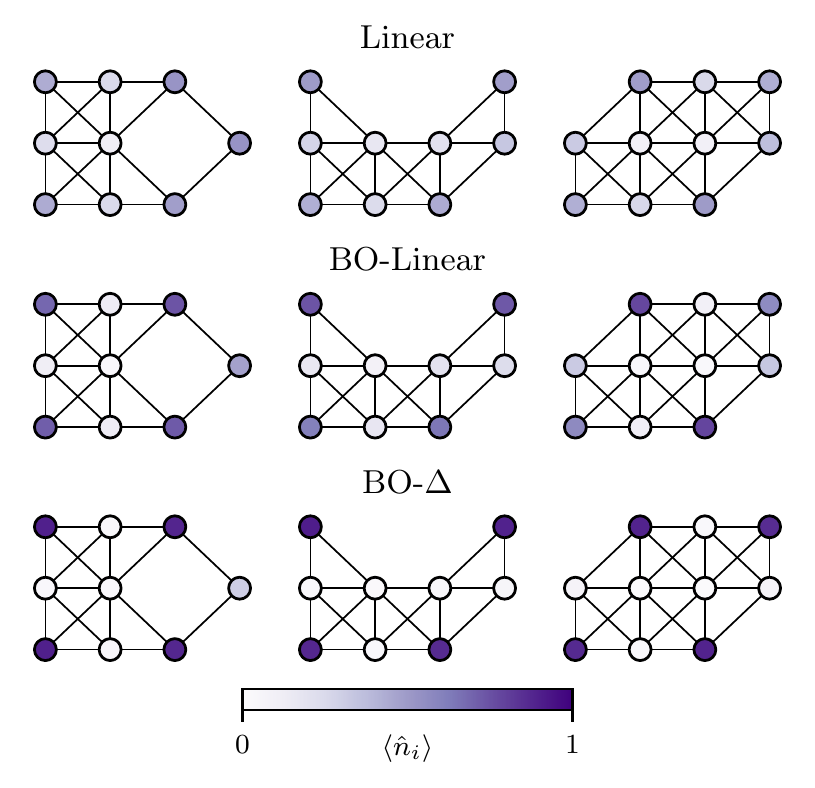}
\caption{
Data from the QuEra Aquila device for the three exemplary instances shown in Fig.~\ref{fig:smallgraphs-schematic}. Here, color denotes the probability of the node being assigned to be in the independent set (purple indicates probability of $1$, see colorbar), for the different protocols described in Section~\ref{sec:results-mis}.
Notice how an improved matching of the colors with the reference schematics in Fig.~\ref{fig:smallgraphs-schematic} corresponds to an improvement in $P_{\mathrm{MIS}}$, as shown in Fig.~\ref{fig:smallgraphs-lineplot}. 
Because the ground state of all of these graphs is unique, disagreement with the ground truth (as shown in Fig.~\ref{fig:smallgraphs-schematic}) stems from experimental noise or from excitations into higher energy states.
}
\label{fig:gridgraphs-colorplot}
\end{figure}
\begin{figure*}[htb]
\subfloat[]{%
    \hspace{-0.05cm}
  \includegraphics{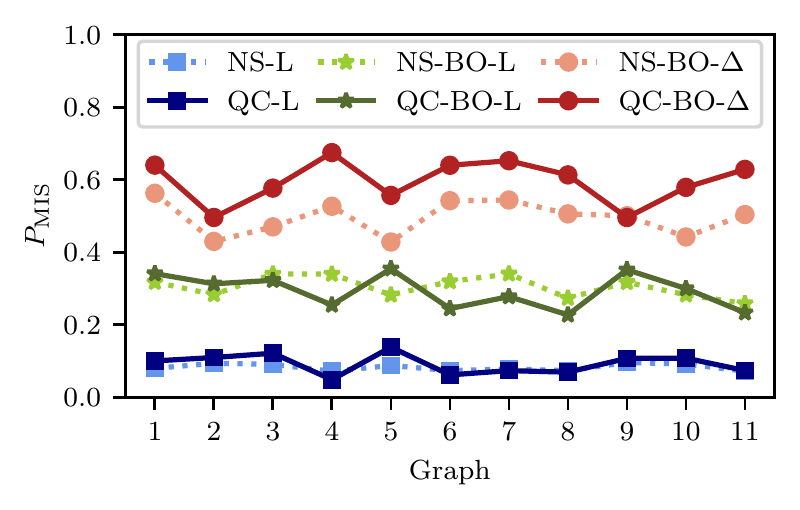}%
  \label{fig:smallgraphs-lineplot}
}
\quad
\subfloat[]{%
  \includegraphics{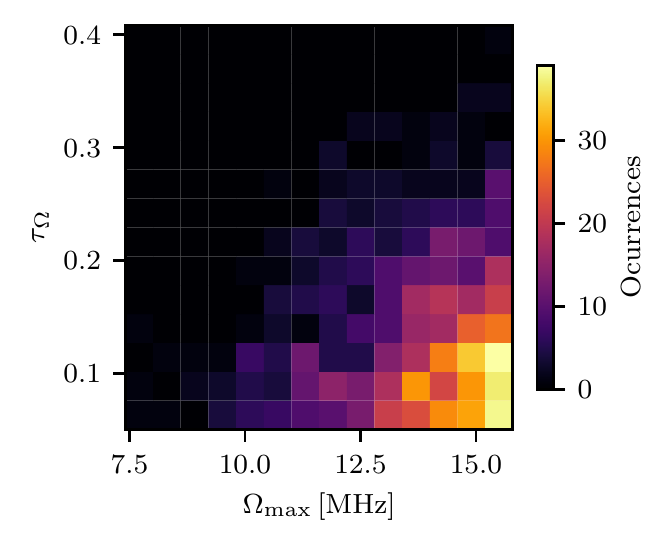}%
  \label{fig:mis-parameters-hist}
}
\caption{(a) Probability of measuring the MIS configuration for the different schedules considered in this manuscript, obtained from classical numerical simulations (NS) or real quantum hardware (QC). The schedule labels L, BO-L, and BO-$\Delta$ correspond to the linear, BO linear, and BO detuning schedules discussed in Sec.~\ref{sec:results-mis}, respectively.
(b) A 2D-histogram of obtained optimal $\Omega$ parameters found by BO for the BO linear schedule for the small graphs. The optimal parameters concentrate at small values of $\tauomega$ and the maximal allowed $\Omega_{\max}$. Here $\tauomega$ is plotted in units of $\tfinal$.
}
\label{fig:app-mis}
\end{figure*}

In this section, we briefly describe the graph generation process and provide additional results on BO-designed schedules for solving MIS on a neutral atom quantum computer. For both the small graphs in Fig.~\ref{fig:smallgraphs-schematic} and the more complex instances in Fig.~\ref{fig:hardgraphs}, the graphs are generated by filling an underlying square lattice with added diagonal connectivity.
These correspond to UDGs with $\sqrt{2}a<R_{\mathrm{d}}<2a$, where $R_{\mathrm{d}}$ is the radius parameter of the UDG and $a$ is the lattice constant.

In the case of the small graphs, the underlying lattice is $4\times 3$ sites. 
We generate all of the $\binom{12}{9} + \binom{12}{10}=286$ graphs with $9$ and $10$ nodes and filter for the ones that have a unique MIS (of which $11$ are mutually non-isomorphic), to simplify our analysis. For these graphs, we then design adiabatic protocols in classical simulations and deploy them on the QuEra Aquila device (see Fig.~\ref{fig:smallgraphs}, Fig.~\ref{fig:gridgraphs-colorplot} and Fig.~\ref{fig:app-mis}). Because the register on the QuEra Aquila device is large enough to accommodate many copies of the small graphs in parallel, and because the parameters of the adiabatic protocol are global (i.e., the same for every atom in the register), we deployed the runs with $12$ copies of the graphs in the register. This parallelization allows us to effectively perform more runs, as a single measurement effectively yields $12$ shots. These shots are additionally postselected on the correct initialization of the corresponding atoms.

For the more complex graphs, the underlying lattice comprises $14\times 14$ sites. We generate graphs with nodes at approximately 70\% of the sites, resulting in graphs with $137$ nodes, and filter for the instances that have large hardness parameters $\mathcal{HP}$ using the \texttt{GenericTensorNetworks} Julia library~\cite{liu_computing_2022}. To limit the number of shots required for the optimization of the schedules for hard graphs, no postselection on the correct initialization of the device is performed.

The parameters for the linear parametrization are chosen as follows: $\deltainitial = \SI{-30}{\mega\hertz}$, $\deltafinal=\SI{60}{\mega\hertz}$, $\Omega_{\max}=\SI{9}{\mega\hertz}$ and $\tau_{\Omega}=0.1\tfinal$. These values were chosen because they comply with the experimental platform and ensure that the ground state of the Rydberg Hamiltonian in Eq.~\eqref{eq:rydberg-hamiltonian} matches the solution to the MIS problem, as discussed in Sec.~\ref{sec:mis-rydberg}).
These parameters are also used as the initial guess supplied to the Bayesian optimizer.

\end{document}